\documentclass[aps,reprint,longbibliography, prb]{revtex4-2}
\usepackage{graphicx}
\usepackage{amsmath, amsfonts}
\usepackage{bbm}
\usepackage{bm}
\usepackage[dvipsnames]{xcolor}
\usepackage{hyperref}
\hypersetup{unicode=true, colorlinks=true, citecolor={blue!80!black}, urlcolor={blue!50!black}, linkcolor = {blue!80!black}}
\usepackage{microtype}
\usepackage[normalem]{ulem}
\usepackage{array}
\usepackage{multirow}

\begin{document}

\title{Microwave response of an Andreev bound state}
\author{Pavel D.~Kurilovich}
\author{Vladislav D.~Kurilovich}
\author{Valla Fatemi}
\author{Michel H.~Devoret}
\author{Leonid I.~Glazman}
\affiliation{Departments of Physics and Applied Physics, Yale University, New Haven, CT 06520, USA}

\date{\today}
\begin{abstract}
We develop a theory for the dynamics of an Andreev bound state (ABS) hosted by a weak link of finite length for which charging effects are important. We derive the linear response of both the current through the link and charge accumulated in it with respect to the phase and gate voltage biases. The resulting matrix encapsulates the spectroscopic properties of a weak link embedded in a microwave resonator. In the low-frequency limit, we obtain the response functions analytically using an effective low-energy Hamiltonian, which we derive. This Hamiltonian minimally accounts for Coulomb interaction and is suitable for a phenomenological description of a weak link having a finite length.
\end{abstract}

\maketitle
\section{Introduction}
Andreev bound states (ABSs) constitute key elements of the microscopic picture of the Josephson effect \cite{houten1991, beenakker1991, furusaki1991}. In a conventional superconducting tunnel junction it is difficult to isolate a single ABS because the Josephson supercurrent is mediated by a large number of shallow ABSs. The situation is different in superconducting weak links based on atomic contacts or  semiconducting nanowires, which recently emerged as a versatile platform for exploring different facets of mesoscopic superconductivity.
In these systems, in contrast to tunnel junctions,  an appreciable supercurrent may be carried by one or a few ABSs stemming from a small number of highly transparent transport channels \cite{bretheau2013, woerkom2017, krogstrup2017, marcus2017}.

The advent of circuit quantum electrodynamics (cQED) brought new experimental capabilities for investigating ABS physics. By coupling the weak link to a microwave resonator, experiments finely resolved separate ABSs and probed their spectrum in various limits \cite{janvier2015, delange2015, hays2018, tosi2019, hays2020, metzger2021}. Time-resolved access to the system provided by cQED also made it possible to use ABSs as qubits. In particular, qubits composed of the occupation of an ABS by zero or two Bogoliubov quasiparticles were implemented in \cite{janvier2015, hays2018}. The operation of such qubits was limited by quasiparticle poisoning: the ABS occasionally trapped a single unpaired quasiparticle, making the qubit leave the computational manifold. Later experiments with semiconducting nanowires showed that the spin of a trapped quasiparticle can also be used as a qubit basis \cite{hays2020,hays2021}.

The manipulation and readout of Andreev qubits rely on interaction with microwave-frequency modes and radiation.
A simplest model that describes the microwave properties of an ABS is that of a short, highly transparent junction with a single transport channel \cite{feigelman1999, zazunov2003, zazunov2005, kos2013}. While this model often works well for atomic contacts, it is insufficient to adequately describe crucial features of ABSs in nanowire devices. First, spectroscopic measurements show that ABSs in nanowire weak links are often situated well within the superconducting gap at any phase bias applied to the junction \cite{woerkom2017, tosi2019, hays2020}. This contrasts the ABS behaviour in a short junction, where the ABS necessarily merges with the edge of the quasiparticle continuum at zero phase bias. Second, ABSs in the experiments \cite{delange2015, larsen2015, hays2018, tosi2019} were sensitive to the gate voltage, pointing to a finite length of the weak links. 
By the same token, properties of nanowire devices are sensitive to charging~\cite{albrecht2016, deng2016}.
Last but not least, a ``poisoned'' ABS hosting a single quasiparticle might carry supercurrent through a weak link of a finite length~\cite{vandam2006}. This aspect is also not present within the short junction model~\cite{beenakker1991}. 

All of the above simplifications of the short junction model come from neglecting the dwell time of a quasiparticle in the junction region.
Usually, the finite dwell time is accounted for by considering microscopic models in which the length of the weak link is comparable to the superconducting coherence length. While such models add an additional realistic aspect for describing the microwave properties of the weak link \cite{tosi2019, metzger2021}, they suffer from being analytically intractable. This complexity often obscures the salient physics of the system. Moreover, Coulomb interaction in the weak link remains unaccounted for in these models. Is it possible to construct an analytically tractable model for describing the microwave properties of a finite-length weak link that would take into account both a finite dwell time and Coulomb interaction?

\begin{figure}[t]
  \begin{center}
    \includegraphics[scale=1]{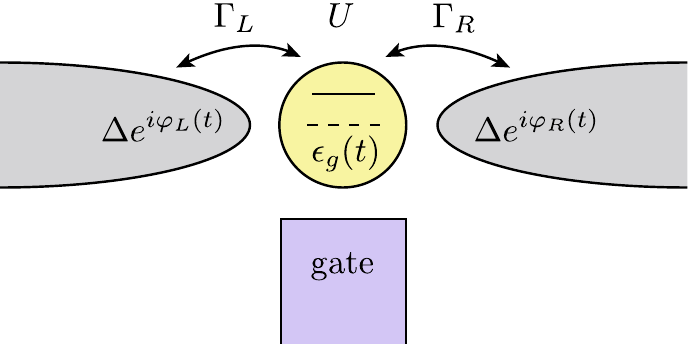}
    \caption{Schematic for a minimal model of a finite-length weak link between two superconductors. A single fermionic level is tunnel coupled to two superconducting leads. The tunneling rates are $\Gamma_L$ and $\Gamma_R$ for the left and right leads, respectively. $\Delta$ is the superconducting order parameter in the leads, $U$ is the strength of the on-site Coulomb interaction. Adjacent gate controls the energy of the level, $\epsilon_g(t)$. Phases of order parameter in the leads are $\varphi_L(t)$ and $\varphi_R(t)$.
    \label{fig:setup}}
  \end{center}
\end{figure}

Here we answer this question affirmatively and calculate the microwave response of a finite-length weak link in a simple phenomenological model that accounts for the dwell time and Coulomb interaction. To build up the model, we assume that the level spacing in an isolated weak link region is large compared to the superconducting order parameter in the leads, $\delta\varepsilon\gg\Delta$. In that case, there is only a single ABS in the system. The finite dwell time, $t_\mathrm{dw}$, introduces the energy scale $\Gamma\equiv \hbar/t_\mathrm{dw}$ which can be interpreted as a normal-state linewidth of levels in the weak link. We consider the case in which this scale may be comparable to $\Delta$. The electrodynamics of the ABS in this regime can be captured by representing the weak link as a single-level quantum dot coupled to two superconducting leads by tunnel junctions \cite{beenakker1992, yeyati2011} (see Fig.~\ref{fig:setup}). One may view such a setting as a generalization \cite{glazman1989}
of the Anderson impurity model \cite{anderson61} with two superconducting reservoirs.
Due to the proximity effect, the level in the dot turns into an ABS whose energy depends not only on the phase difference across the weak link but also on the voltage applied to an adjacent gate. Keeping in mind quasiparticle poisoning \cite{janvier2015, hays2018}, we find the microwave response in states with both even and odd fermion parity.

\subsection*{Summary of results}
Below we summarize our main results. After writing down the fundamental aspects of our model [Sec.~\ref{sec:model}], in Sec.~\ref{sec:spectrum} we find the ABS energy spectrum as a function of its occupancy with quasiparticles and applied phase and gate voltage biases.
The main parameters that control these dependencies are the rates of tunneling between the dot and the leads $\Gamma_L$, $\Gamma_R$ (expressed in energy units), the superconducting order parameter $\Delta$, and the on-site Coulomb repulsion $U$. Our results for the energies are valid at arbitrary ratio $\Gamma / \Delta$ (where $\Gamma = \Gamma_L + \Gamma_R$) provided that Coulomb interaction is weak, $U \ll \Delta + \Gamma$, so that the Kondo effect does not develop \footnote{We note that the requirement of weak interaction is usually at odds with the requirement of the large level spacing. However, the two requirements can be simultaneously met if the interaction is screened by the leads or by the gate.}.

In an experimentally relevant limit $\Gamma\sim\Delta$ \cite{marcus2020, kou2020, fatemi2021}, the continuum of occupied levels outside of the gap gives a substantial contribution to the phase- and gate voltage-dependencies of the energy \cite{beenakker1992}.
In the odd states, the energy is fully determined by the continuum contribution. This contribution reaches its minimum at $\varphi = \pi$; thus, the weak link is a $\pi$-junction when a single quasiparticle occupies the ABS. 
While the latter fact is well-known for strongly-interacting Anderson impurities (where the odd state is the ground state), to our knowledge, the $\pi$-junction behavior was not appreciated for a weakly-interacting ABS poisoned by a quasiparticle.
An example of the phase and gate voltage dependencies of the discrete energy levels is presented in Fig.~\ref{fig:energy}.

In the even fermion parity sector the ABS forms an Andreev pair qubit \cite{zazunov2003}. For small drive frequencies, $\hbar\omega \ll \Delta$, the dynamics of this qubit can be described with the help of an effective low-energy Hamiltonian presented in Sec.~\ref{sec:low-energy} [see Eq.~\eqref{eq:low-en2}]. Our Hamiltonian smoothly interpolates between the Hamiltonian of a quantum dot weakly coupled to the superconducting leads ($\Gamma \ll \Delta$) \cite{belzig2020, oriekhov2021} and the Hamiltonian of a short tunnel junction ($\Gamma \gg \Delta$) \cite{feigelman1999, zazunov2003, zazunov2005}.

Next, in Sec.~\ref{sec:lin-res} we investigate the linear electrodynamic response of a Josephson weak link containing an ABS.
Due to the finite length of the weak link a non-zero charge can be accumulated in the junction region (in contrast to the atomic point contact). 
Therefore, the response function has a structure of a $2 \times 2$ matrix: current through the weak link and charge on it respond to the phase and gate biases. We compute this matrix for many-body states with the different number of quasiparticles at the ABS. If the drive frequency is small, the response matrix describes the quasi-static characteristics of the weak link such as the inverse inductance and quantum capacitance. At higher frequencies the response exhibits a resonance, if the occupancy of the ABS is even [see Fig.~\ref{fig:response}].

The response functions of the ABS are sensitive to the presence of Coulomb interaction. The strength of the interaction $U$ can be deduced from the response functions measured in states with different occupation of the ABS. In Sec.~\ref{sec:asym} we identify a particularly convenient quantity which gives a direct access to the magnitude of interaction, $\chi_\mathrm{asym} = \frac{\chi[0]+\chi[2]}{2}-\chi[1]$ (where the arguments in the square brackets correspond to the occupation of the ABS with quasiparticles). This quantity is convenient because $\chi_\mathrm{asym}$ vanishes in the absence of interaction, i.e. an occupation rule $\frac{\chi[0]+\chi[2]}{2}=\chi[1]$ is satisfied at $U = 0$.

The response functions can be measured experimentally in a circuit quantum electrodynamics (cQED) architecture by coupling the weak link to a microwave resonator and measuring the dispersive shift of the latter. We demonstrate how specific components of the matrix response function $\chi$ may be singled out by tailoring the geometry of the resonator [see Sec.~\ref{sec:cqed}].

Our theory thus provides a guide for analyzing measurements of the microwave response of finite-length nanowire weak links. 

\section{Model\label{sec:model}}
{We consider a weak-link between two superconducting leads with a solitary ABS. We assume that the linewidth for a quasiparticle in the weak link, $\Gamma\equiv \hbar/t_\mathrm{dw}$, might be comparable to $\Delta$ while the level spacing of the link $\delta\varepsilon \gg \Delta$.
It is possible to model such a weak link as a quantum dot tunnel-coupled to two superconducting leads (see Fig.~\ref{fig:setup} for the schematics).} The Hamiltonian of this system reads
\begin{equation}
\label{eq:model}
H[\epsilon_g(t), \varphi_{i}(t)] =\sum_{i=L,R}\Bigl[ H_{i} + H_{T,i}[\varphi_{i}(t)]\Bigr]+H_{d}[\epsilon_g (t)].
\end{equation}
The Hamiltonian of the lead $i$ ($i=L/R$ denotes the left/right lead, respectively) is:
\begin{equation}
H_i=\int d \mathbf{r} \Bigl[\sum_\sigma\psi_{\sigma,i}^\dagger(\mathbf r)\hat{\xi} \psi_{\sigma,i}(\mathbf r)+\Delta (\psi_{\downarrow,i}(\mathbf r) \psi_{\uparrow,i}(\mathbf r)+\mathrm{h.c.})\Bigr].
\end{equation}
Here $\psi_{\sigma, i}(\mathbf r)$ and $\psi_{\sigma,i}^\dagger( \mathbf r)$ are the annihilation and creation operators of an electron with spin $\sigma =\,\uparrow$ or $\downarrow$ in the lead $i$,
$\hat{\xi}$ is the operator of kinetic energy with the respect to the Fermi level, and
$\Delta$ is the $\mathrm{s}$-wave superconducting gap, identical in the two leads. We work in a gauge in which the superconducting phases of the leads are attached to the tunneling amplitudes, cf.~Eq.~\eqref{eq:tunneling}. Thus we assume $\Delta > 0$.

The ``quantum dot'' is described by
\begin{equation}
\label{eq:res-lev}
H_{d}[\epsilon_g (t)] =\sum_{\sigma}\epsilon_g (t) d_{\sigma}^{\dagger}d_{\sigma} + U (d_\uparrow^\dagger d_\uparrow-1/2)(d_\downarrow^\dagger d_\downarrow - 1/2),
\end{equation}
where $\epsilon_g(t)$ is determined by the applied gate voltage, $\epsilon_g(t)= - e V_g(t)$ ($e>0$). The latter can have both a static and a dynamic part, $V_g(t) = V_g + \delta V_g(t)$, where $V_g$ controls the energy of the fermionic level at the dot and $\delta V_g(t)$ describes the external driving. $U>0$ is the energy of Coulomb interaction at the level. Throughout the manuscript we assume that the Coulomb interaction is weak, $U \ll \Delta + \Gamma$. The reference point for the gate voltage is chosen such that $\epsilon_g = 0$ is the charge neutrality point.

The remaining term in Eq.~\eqref{eq:model} describes the tunneling between the dot and the lead $i$ and is given by
\begin{equation}
\label{eq:tunneling}
H_{T,i}[\varphi_{i}(t)]=t_{i}\sum_\sigma (e^{\frac{i}{2}\varphi_{i}(t)}d^{\dagger}_{\sigma}\psi_{\sigma,i}(0)+\mathrm{h.c.}),
\end{equation}
where $t_i$ is the tunneling amplitude. The superconducting phase $\varphi_i(t)$ has static and dynamic parts, $\varphi_i(t) = \varphi_i + \delta\varphi_i(t)$. The dynamic component of each phase is related to the dynamic part of the voltage applied to the respective lead, $\delta V_i(t)$, via the Josephson relation $\phi_0\partial_t\delta\varphi_i(t)=\delta V_i(t)$, where $\phi_0 = \hbar/2e$ is the reduced flux quantum. It is convenient to characterize the tunneling between the dot and the leads by the corresponding tunneling rates $\Gamma_i = \pi \nu t_i^2$, where $\nu$ is the normal-state density of states at the Fermi level in the leads (per spin species). We also introduce the total tunneling rate $\Gamma$ and the difference of tunneling rates $\delta\Gamma$:
\begin{equation}
\Gamma = \Gamma_L + \Gamma_R,\quad\quad \delta \Gamma = \Gamma_L - \Gamma_R.
\end{equation}
Throughout our work we assume for simplicity that the rates $\Gamma_L$ and $\Gamma_R$ do not depend on the applied gate voltage.

In our model, we assume that the capacitance between the dot and the gate is much larger than the capacitances between the dot and the superconducting leads. This assumption is justified if the gate is located sufficiently close to the weak link. We also neglect the capacitance between the dot and the ground. In this case, voltage $V_g(t)$ applied to the gate is equivalent to voltage $-V_g(t)$ with respect to the ground applied simultaneously to both leads, as these two situations differ by an overall shift of energy. Using this freedom we choose $\varphi_L(t)=-\varphi_R(t)=\varphi(t)/2$. We discuss how our theory is modified at an arbitrary ratio between the capacitances to the gate and to the leads in Appendix \ref{app:cap}.

Finally, we introduce the operators of the charge at the dot, $\hat{Q} = -e\sum_\sigma d^\dagger_\sigma d_\sigma$, and of the current flowing through the weak link, $\hat{I}= -\frac{e}{2}\frac{d}{dt} (N_R-N_L)$, where $N_i$ is operator of the number of electrons in lead $i$. We note that $\frac{d}{dt}\hat{Q} = e\frac{d}{dt}(N_L+N_R)$ due to charge conservation.

\section{Energy spectrum}
\label{sec:spectrum}

\begin{figure*}[t]
  \begin{center}
    \includegraphics[scale=1]{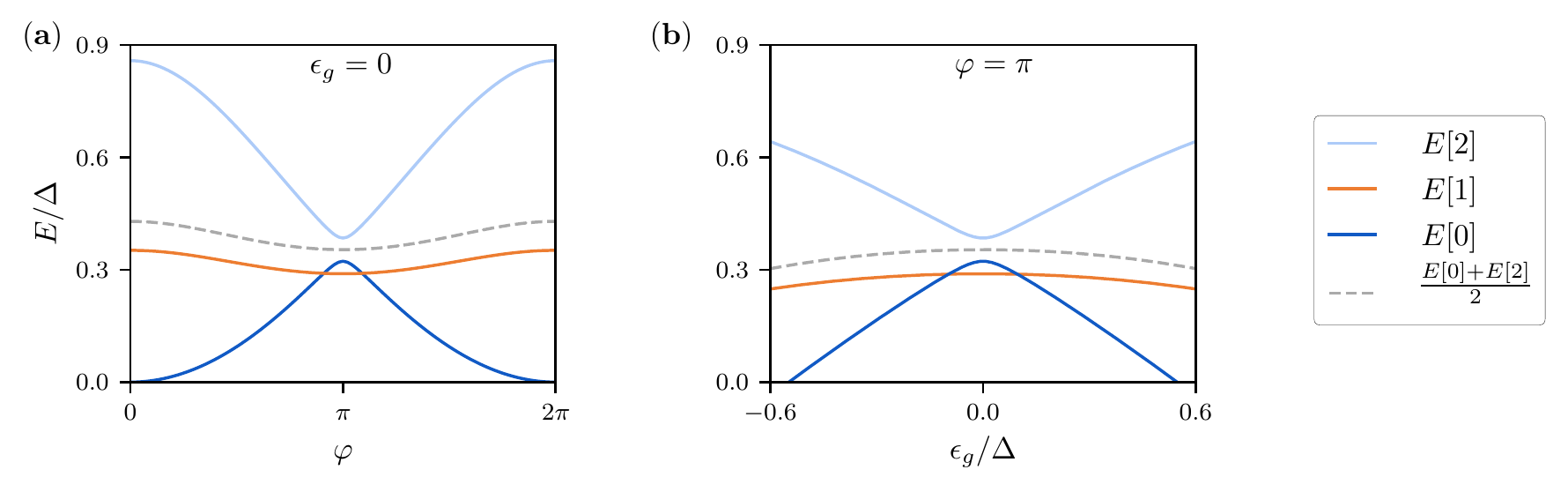}
    \caption{Energies $E[n]$ of states with $n$ quasiparticles at the Andreev bound state as functions of $\varphi$ [panel (a)] and $\epsilon_g = - e V_g$ [panel (b)]; the energies are calculated with respect to $E[0]$ evaluated at $\varphi = 0$ and $\epsilon_g = 0$. The plots are produced using Eqs.~\eqref{eq:E02}-\eqref{eq:Aij}, \eqref{eq:even}, \eqref{eq:odd} with
    $\Gamma_L = 0.3 \Delta$, $\Gamma_R = 0.35 \Delta$, $U = 0.35 \Delta$. The phase dependence in panel (a) is plotted for $\epsilon_g = 0$. The gate voltage dependence in panel (b) is plotted for $\varphi = \pi$; note that $\epsilon_g$ is computed with respect to the charge-degeneracy point  $\epsilon_g = 0$. Dashed line shows the half-sum between energies $E[0]$ and $E[2]$. The fact that the dashed line does not coincide with $E[1]$ is due to the presence of Coulomb interaction at the resonant level; the difference between $(E[0]+E[2])/2$ and $E[1]$ is given by $E_\mathrm{asym}$, see Eq.~\eqref{eq:asym-vs-U}. Note that close to $\varphi = \pi$ and $\epsilon_g=0$ even a weak Coulomb interaction can render the ground state of the system odd in the electron number.
    \label{fig:energy}} 
  \end{center}
\end{figure*}

We initially assume that the phase bias $\varphi$ and gate voltage~$V_g$ are static and study the many-body energy spectrum of model defined by Eqs.~\eqref{eq:model}-\eqref{eq:tunneling}. For weak Coulomb interaction there are four discrete energy levels in the spectrum that are separated from the many-body continuum. We refer to these states as $|0\rangle$, $|1_\uparrow\rangle$, $|1_\downarrow\rangle$, and $|2\rangle$. The four states correspond to a different number of Bogoliubov quasiparticles occupying the ABS:
zero, one (with spin up or down), or two, respectively \footnote{We define quasiparticles with the respect to the lowest energy even state. For weak enough interaction this state is the ground state of the system. However, this might not be the case for stronger interactions in some domain of control parameters, see later discussion and Fig.~\ref{fig:energy}. Still, for uniformity of narrative we always refer to the lowest energy even state as the one with no quasiparticles.}. We denote the energies of the states as $E[0]$, $E[1]$, $E[2]$ (states $|1_\uparrow\rangle$ and $|1_\downarrow\rangle$ are spin-degenerate). The energies of the levels, $E[n]\equiv E[n, \epsilon_g, \varphi]$, depend on $\epsilon_g = - e V_g$ and phase bias $\varphi$.

We start with a detailed description of the many-body spectrum in the absence of Coulomb interaction. We then take $U\neq 0$ into account perturbatively. At $U = 0$ the energies of the discrete levels can be expressed as (see Appendix \ref{app:mb_spectrum}) \footnote{Strictly speaking, there is also a contribution $\epsilon_g$ to energy of all states. However, it does not affect the physical properties of the system --- such as the linear response functions --- and henceforth we omit it throughout the text.}
\begin{equation}
\label{eq:unpert}
E_\mathrm{0}[n] = E_\mathrm{cont} + (n - 1) E_\mathrm{A,0},
\end{equation}
where the subscript $0$ indicates that $U = 0$. Here, $E_\mathrm{cont}$ is the energy associated with the continuum states (see later discussion). $E_{\mathrm{A,0}} > 0$ is the energy of the ABS which can be found by solving characteristic equation
\begin{equation}\label{eq:abs_unp_energy}
\det G_{dd}^{-1}(\varepsilon) = 0
\end{equation}
at $0 \leq \varepsilon < \Delta$. Here $G_{dd}(\varepsilon)$ is the Green's function of the dot at $U = 0$ (see Appendix \ref{app:gf} for the derivation):
\begin{equation}
\label{eq:G-inv}
G^{-1}_{dd}(\varepsilon) = \frac{\varepsilon}{Z(\varepsilon)} - \epsilon_g\tau_z - \sum_{i=L,R}\frac{\Delta\Gamma_i e^{\frac{i}{2} \tau_z \varphi_i} \tau_x e^{-\frac{i}{2} \tau_z \varphi_i}}{\sqrt{\Delta^2 - \varepsilon^2}}.
\end{equation}
In this expression, $\tau_{x,y,z}$ are Pauli matrices in the Nambu space,
\begin{equation}
\label{eq:Z}
\frac{1}{Z(\varepsilon)} = 1 + \frac{\Gamma}{\sqrt{\Delta^2 - \varepsilon^2}},   
\end{equation}
and $\varphi_L = -\varphi_R = \varphi/2$.

The energy of the ABS, $E_\mathrm{A,0}$, can be written analytically when the parameters of the system are tuned such that $E_{\mathrm{A,0}}\ll \Delta$. The latter condition is satisfied, at any phase bias, if the level at the dot is located close to the Fermi energy and is weakly coupled to the leads, $\Gamma, |\epsilon_g| \ll \Delta$. It is also satisfied when the coupling to the leads is strong, $\Gamma \gtrsim \Delta$, provided $|\delta\Gamma|, |\epsilon_g| \ll \Gamma$ and $|\varphi-\pi|\ll 1$. In either case we obtain an approximate solution
\begin{equation}
    \label{eq:low-en-abs}
    E_\mathrm{A,0} = \frac{\Delta}{\Delta + \Gamma} \sqrt{\epsilon_g^2 + |\gamma|^2},\:\: \gamma = \Gamma \cos\frac{\varphi}{2} + i \delta\Gamma \sin\frac{\varphi}{2}.
\end{equation}
The prefactor $\Delta/(\Delta + \Gamma)$ characterizes the extent to which the wave-function of the ABS is localized at the quantum dot. If the tunneling between the leads and the dot is weak, $\Gamma \ll \Delta$, the wave-function is predominantly localized at the dot and $\Delta/(\Delta + \Gamma) \approx 1$. In this regime, Eq.~\eqref{eq:low-en-abs} reproduces the known result for the ABS energy $E_\mathrm{A,0}$ \cite{wendin1996, devyatov1997}. For stronger tunneling, $\Gamma \gtrsim \Delta$, prefactor $\Delta/(\Delta + \Gamma) < 1$ due to the spreading of the wave-function from the dot into the leads. If $\Gamma \gg \Delta$ the support of the wave-function is mostly in the leads.  In this case, the system is essentially equivalent to a short junction. In particular, after approximating the prefactor in Eq.~\eqref{eq:low-en-abs} as $\Delta/(\Delta+\Gamma) \approx \Delta/\Gamma$ and neglecting $\epsilon_g$ under the square root, Eq.~\eqref{eq:low-en-abs} reproduces the energy spectrum of a short junction \cite{houten1991} with the reflection amplitude $r = \delta\Gamma/\Gamma$.

Note that, regardless of the ratio between $\Gamma$ and $\Delta$, the approximate functional form of the dependence of $E_\mathrm{A,0}$ on $\varphi$ in Eq.~\eqref{eq:low-en-abs} is similar to that for a short junction, 
\begin{equation}
\label{eq:func-form}
E_\mathrm{A,0} = \delta \sqrt{1-\tau \sin^2(\varphi/2)}.
\end{equation}
There is, however, an important difference: in our model the level is generally detached from the superconducting gap at all phase biases, $\delta < \Delta$, whereas for the short junction $\delta = \Delta$ and $E_{\rm A,0}$ therefore reaches $\Delta$ at $\varphi = 0$.

{In Eq.~\eqref{eq:unpert}, $E_\mathrm{cont}$ is the energy associated with the continuum~\cite{beenakker1992} of filled single-particle states at $\varepsilon < -\Delta$,
\begin{equation}
    \label{eq:cont}
    E_\mathrm{cont} = \int_{-\infty}^{-\Delta} \frac{d\varepsilon}{2\pi i} \ln\det\left[G_{dd}^A(\varepsilon)\left[G_{dd}^R(\varepsilon)\right]^{-1}\right],
\end{equation}
where $G^{R/A}_{dd}(\varepsilon) = G_{dd}(\varepsilon \pm i0)$ \footnote{In Eq.~\eqref{eq:cont} we omitted the contribution $\propto E_F N$, where $E_F$ is the Fermi energy in the leads and $N$ is the number of electrons in the leads. This contribution does not depend on phase and gate voltage and thus does not affect response functions.}.
This energy depends on the gate voltage and phase due to the coupling between the leads and the dot,  $E_\mathrm{cont}\equiv E_\mathrm{cont}(\epsilon_g,\varphi)$.}
{The integral in Eq.~\eqref{eq:cont} is divergent at the lower limit.
However, this divergence does not influence the observables. Indeed, the integral for the difference $E_\mathrm{cont}(\epsilon_g,\varphi) - E_\mathrm{cont}(0,0)$ converges at $|\varepsilon| \sim \Delta$ whereas the divergent contribution $E_\mathrm{cont}(0,0)$ does not depend on $\epsilon_g$ and $\varphi$.} Energy ${E}_\mathrm{cont}$ can be found analytically in the limit of weak coupling, $\Gamma \ll \sqrt{\Delta^2 - \epsilon_g^2}$,
\begin{align}
    &E_\mathrm{cont}(\epsilon_g, \varphi)-E_\mathrm{cont}(0,0) =\notag\\
    &= -\frac{2}{\pi} \Gamma \frac{\epsilon_g \arcsin(\epsilon_g/\Delta)}{\sqrt{\Delta^2-\epsilon_g^2}} - \frac{4\Delta \Gamma_R \Gamma_L \sin^2(\varphi/2) }{\Delta^2-\epsilon_g^2}.
\label{eq:cont-weak-coupling}
\end{align}
By extrapolating Eq.~\eqref{eq:cont-weak-coupling} to $\Gamma \sim \Delta$ we observe that in this regime $E_\mathrm{cont}(\epsilon_g, \varphi)-E_\mathrm{cont}(0,0)$ is of the same order as $E_{\mathrm{A,0}}$ [cf.~Eq.~\eqref{eq:low-en-abs} and Eq.~\eqref{eq:cont-weak-coupling}]. Thus, $E_\mathrm{cont}$ may strongly contribute to the observable properties of the system, such as the linear response functions [see Sec.~\ref{sec:lin-res} for discussion of this].

We note that in the absence of Coulomb interaction the phase-dependence of the energy of the odd states, $E_0[1]$, is determined solely by $E_\mathrm{cont}$ [see Eq.~\eqref{eq:unpert}]. Equation \eqref{eq:cont-weak-coupling} thus demonstrates that the system realizes a $\pi$-junction in states $|1_{0,\uparrow}\rangle$, $|1_{0,\downarrow}\rangle$, \textit{i.e.}, $E_0[1]$ is minimal at $\varphi = \pi$.

To conclude the discussion of the non-interacting case, we note that at $U = 0$ the ground state of the system always has even fermion number parity as follows directly from Eq.~\eqref{eq:unpert}.

\begin{figure}[t]
  \begin{center}
    \hspace{-20pt}\includegraphics[scale=1]{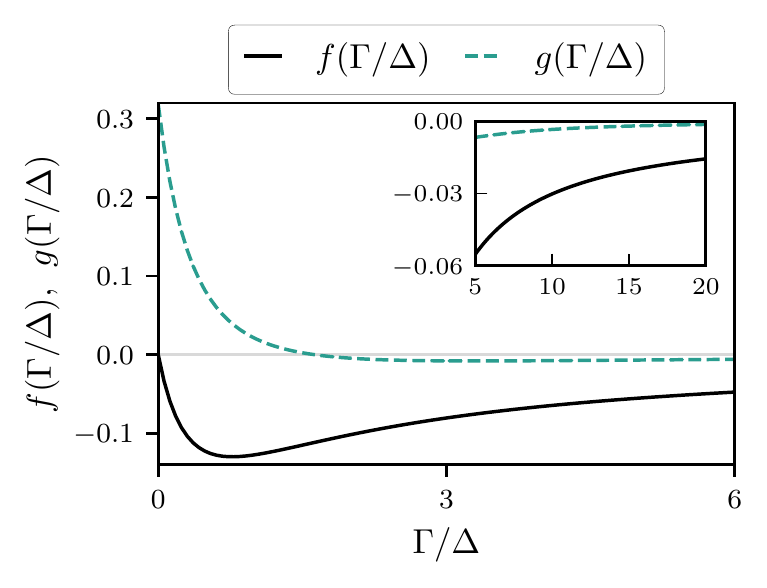}
    \caption{Functions $f(\Gamma/\Delta)$ and $g(\Gamma/\Delta)$ that determine how the energy of the ABS, $E_\mathrm{A}$, is renormalized due to a weak Coulomb interaction [see Eq.~\eqref{eq:renorm}]. As it follows from Eqs.~\eqref{eq:f} and \eqref{eq:g}, at $\Gamma\ll\Delta$ we may approximate $f = - \frac{2}{\pi}\frac{\Gamma}{\Delta}$ and $g = \frac{1}{\pi} - \frac{\Gamma}{\Delta}$. In the opposite limit, $\Gamma \gg \Delta$, we find $f= -\frac{1}{\pi}\frac{\Delta}{\Gamma}$ and $g=-\frac{1}{\pi}\left(\frac{\Delta}{\Gamma}\right)^2\ln\left(\frac{2\Gamma}{e^2\Delta}\right)$.
    \label{fig:f-and-g}}
  \end{center}
\end{figure}

Next, we apply the first order perturbation theory in $U$ to approximately find the energies of the discrete states in the presence of Coulomb interaction. We start with the states in the even fermion parity sector, $|0\rangle$ and $|2\rangle$. At $U \neq 0$ it is convenient to parametrize their energies as
\begin{equation}
\label{eq:E02}
E[0] = E_\mathrm{even} - E_\mathrm{A},\quad E[2] = E_\mathrm{even} + E_\mathrm{A}.
\end{equation}
To determine $E_\mathrm{A}$ and $E_\mathrm{even}$, we project the interaction Hamiltonian, $H_\mathrm{int} = U (d_\uparrow^\dagger d_\uparrow - 1/2) (d_\downarrow^\dagger d_\downarrow - 1/2)$, onto the subspace formed by the unperturbed states $|0_0\rangle$ and $|2_0\rangle$, thus constructing a characteristic equation.
The projection is carried out conveniently is the basis of particle and hole states that can be obtained from $|0_0\rangle$ and $|2_0\rangle$ by a proper rotation. We find the following equation for $E_\mathrm{A}$ (see Appendix \ref{app:int} for details of the derivation):
\begin{equation}
\label{eq:long-det}
    \det\left[\varepsilon - \mathcal{H} - U\alpha
\begin{pmatrix}
\frac{A_{pp}-A_{hh}}{2}&A_{ph}\\
A_{ph}^\star&-\frac{A_{pp}-A_{hh}}{2}
\end{pmatrix}
\right] = 0,
\end{equation}
where ${\cal H}$ is related to the non-interacting Green's function, ${\cal H} = E_{\rm A,0} - Z(E_{\rm A,0}) G^{-1}_{dd}(E_{\rm A,0})$ [see Eqs.~\eqref{eq:G-inv}, \eqref{eq:Z}]. Explicitly,
\begin{equation}
    \mathcal{H} = \frac{1}{1+\frac{\Gamma}{\sqrt{\Delta^2 - E_\mathrm{A,0}^2}}}
    \begin{pmatrix}
    \epsilon_g & \frac{\Delta}{\sqrt{\Delta^2 - E_\mathrm{A,0}^2}}\gamma\\
    \frac{\Delta}{\sqrt{\Delta^2 - E_\mathrm{A,0}^2}}\gamma^\star& -\epsilon_g
    \end{pmatrix}
\end{equation}
with $\gamma\equiv\gamma[\varphi]= \Gamma\cos(\varphi/2) + i\delta\Gamma \sin(\varphi/2)$.
Functions $A_{ij}\equiv A_{ij}[\epsilon_g, \varphi]$ are defined as
\begin{equation}
    \label{eq:Aij}
    A_{ij} = - \int_{-\infty}^{-\Delta} \frac{d\varepsilon}{2 \pi i} \left[G_{dd}^R(\varepsilon) - G_{dd}^A(\varepsilon)\right]_{ij},
\end{equation}
where $i,j = p/h$ are the Nambu indices that correspond to particles and holes, respectively. The parameter $\alpha \equiv \alpha[\epsilon_g, \varphi] $ in Eq.~\eqref{eq:long-det} is related to the matrix $A$ via
\begin{equation}\label{eq:alpha_int}
\alpha = 1 - \mathrm{tr}A.
\end{equation}
It can be explicitly expressed through the bound state energy $E_\mathrm{A,0}$:
\begin{equation}
\label{eq:alpha}
    \frac{1}{\alpha} - 1= \frac{\Delta^2}{\Delta^2 - E_{\mathrm{A,0}}^2}\frac{\Gamma + \frac{4\Gamma_R\Gamma_L \sin^2(\varphi/2)}{\sqrt{\Delta^2 - E_{\mathrm{A,0}}^2}}}{\Gamma + \sqrt{\Delta^2 - E_{\mathrm{A,0}}^2}}.
\end{equation}
At arbitrary $\Gamma, \epsilon_g,\varphi$ equations \eqref{eq:long-det}-\eqref{eq:Aij} for $E_\mathrm{A}$ can be analyzed numerically. An explicit approximate solution can be obtained when $E_\mathrm{A,0}\ll\Delta$. In this case, we find
\begin{equation}
    \label{eq:EA}
    E_{\mathrm{A}} = \frac{\Delta}{\Delta + \Gamma}\sqrt{\tilde{\epsilon}_g^2 + |\tilde{\gamma}[\varphi]|^2},
\end{equation}
where
\begin{equation}
\label{eq:renorm}
    \tilde{\epsilon}_g = \left[1+\frac{U}{\Delta}f\right] \epsilon_g,\quad \tilde{\gamma}[\varphi] =  \left[1+\frac{U}{\Delta}g\right] \gamma[\varphi].
\end{equation}
Here, functions $f \equiv f(\Gamma/\Delta)$ and $g \equiv g(\Gamma/\Delta)$ depend only on the total tunneling rate $\Gamma$ and contain no dependence on $\varphi$ and $\epsilon_g$ with the considered precision. These functions describe the renormalization of $\epsilon_g$ and $\gamma[\varphi]$ in the expression for $E_A$ by the Coulomb interaction [cf.~Eqs.~\eqref{eq:low-en-abs} and \eqref{eq:EA}].
Explicit expressions for $f$ and $g$ are cumbersome and so are presented in Appendix \ref{app:int-low-en}.
The dependence of $f$ and $g$ on the ratio $\Gamma/\Delta$ in demonstrated in Fig.~\ref{fig:f-and-g}.

We proceed by calculating energy $E_\mathrm{even}$ in Eq.~\eqref{eq:E02}. At $U = 0$, $E_\mathrm{even} \equiv E_0[1] = E_\mathrm{cont}$, cf.~Eq.~\eqref{eq:unpert}. The first order correction to this expression due to the weak Coulomb interaction is given by $\mathrm{Tr}\,H_\mathrm{int}/2$, where the trace is computed over the unperturbed discrete states in the even sector, $|0_0\rangle$ and $|2_0\rangle$. Practically, it is again more convenient to perform the calculation in the particle-hole basis (see Appendix \ref{app:int}). By finding the trace we obtain
\begin{gather}
    \label{eq:even}
    E_\mathrm{even} = E_0[1] - U\det\left(A - \frac{1}{2}\right) + \frac{U}{2}\alpha^2.
\end{gather}
At arbitrary $\Gamma,\epsilon_g$, and $\varphi$, the energy $E_\mathrm{even}$ can be computed numerically using Eqs.~\eqref{eq:cont}, \eqref{eq:Aij}, \eqref{eq:alpha}. An explicit analytic expression for $E_\mathrm{even}$ can be obtained if $E_\mathrm{A,0}\ll \Delta$. In this case, we may approximate
\begin{equation}
    \label{eq:alphas}
    A_{pp}\approx A_{hh} \approx \frac{1-\alpha}{2},\quad \alpha \approx \frac{\Delta}{\Delta + \Gamma},
\end{equation}
and neglect $|A_{ph}|\ll A_{pp}, A_{hh}$.
When the tunneling between the dot and the leads is strong, $\Gamma/\Delta \gg 1$, the ABS spreads from the dot into the leads, and $\alpha \ll 1$.
This dilutes the effects of interaction. Indeed, using Eq.~\eqref{eq:alphas} in Eq.~\eqref{eq:even} we estimate $E_\mathrm{even}-E_0[1] = U\Delta^2/4\Gamma^2 \ll U$. In the opposite limit, $\Gamma\ll\Delta$, the ABS is localized at the dot: $\alpha \approx 1$ and $E_\mathrm{even}-E_0[1] = U/4$. We note that in the leading approximation in $E_{\rm A,0} / \Delta$ parameters $\alpha$ and $A_{ij}$ are independent of $\varphi$ and $\epsilon_g$ [see Eq.~\eqref{eq:alphas}]. However, this is only an approximation. These dependencies may be explicitly quantified in the weak coupling limit $\Gamma, |\epsilon_g| \ll \Delta$.
However, the results are cumbersome, we present them in Appendix \ref{app:int-weak-coup} [see Eq.~\eqref{eq:even-weak-coup}].

Finally, we calculate the energy of the odd states, $E[1]$. To this end, we note that spin conservation prevents the Coulomb interaction from coupling the unperturbed states $|1_{0,\uparrow}\rangle$ and $|1_{0,\downarrow}\rangle$. Thus, non-degenerate perturbation theory can be used to find the corrections to their energies, $E[1] \approx E_0[1] + \langle 1_{0,\sigma} |H_\mathrm{int}|1_{0,\sigma}\rangle$ (note that $\langle 1_{0,\sigma} |H_\mathrm{int}|1_{0,\sigma}\rangle$ does not depend on $\sigma$). By computing the matrix element (see Appendix \ref{app:int}) we obtain
\begin{gather}
    \label{eq:odd}
    E[1] = E_0[1] - U\det\left(A - \frac{1}{2}\right).
\end{gather}
When $E_\mathrm{A,0}\ll \Delta$ and $\Gamma/\Delta \gg 1$ we obtain $E[1]-E_0[1] \approx -U\Delta^2/4\Gamma^2 \ll U$. For $\Gamma/\Delta \ll 1$ we find $E[1]-E_0[1] \approx -U/4$. The dependence of $E[1] - E_0[1]$ on $\varphi$  and $\epsilon_g$ in the weak coupling limit ($\Gamma, \epsilon_g \ll \Delta$), is presented in Appendix \ref{app:int-weak-coup} [see Eq.~\eqref{eq:odd-weak-coup}].

An example of phase and gate-voltage dependence of energies $E[n]$ obtained numerically with Eqs.~\eqref{eq:E02}-\eqref{eq:Aij}, \eqref{eq:even}, \eqref{eq:odd} is demonstrated in Fig.~\ref{fig:energy}.

A notable feature demonstrated by Fig.~\ref{fig:energy} is the asymmetry between the level spacings $E[2] - E[1]$ and $E[1] - E[0]$. To characterize this asymmetry we introduce the difference
\begin{equation}
    \label{eq:en-asymmetry}
    E_\mathrm{asym} =\frac{E[2]+E[0]}{2}-E[1].
\end{equation}
In the absence of Coulomb interaction $E_\mathrm{asym} = 0$, as follows directly from Eq.~\eqref{eq:unpert}. Thus, the level-spacing asymmetry ($E_\mathrm{asym} \neq 0$) is a direct consequence of Coulomb repulsion at the ABS. This can be illustrated by considering a simple case in which the tunneling between the dot and the leads is turned off ($\Gamma=0$). Then we find $E[1] - E[0] = \epsilon_g - U/2$ and $E[2] - E[1] = \epsilon_g + U/2$ and thus $E_\mathrm{asym} = U/2$. In the presence of tunneling, we use  Eqs.~\eqref{eq:E02}, \eqref{eq:even}, and \eqref{eq:odd} and obtain
\begin{equation}
    \label{eq:asym-vs-U}
    E_\mathrm{asym} = \frac{U}{2}\alpha^2
\end{equation}
with $\alpha$ given by Eq.~\eqref{eq:alpha}. To highlight the asymmetry in Fig.~\ref{fig:energy}, in addition to $E[1]$ (solid orange curve) we present $(E[0] + E[2])/2$ (dashed gray curve). The mismatch between the two curves is determined by $E_\mathrm{asym}$, see Eq.~\eqref{eq:en-asymmetry}. We note that when $E_\mathrm{A,0}\ll\Delta$ equation \eqref{eq:asym-vs-U} for $E_\mathrm{asym}$ can be simplified. Then, using Eq.~\eqref{eq:alphas} we obtain an approximate expression for the asymmetry,
\begin{equation}
    \label{eq:asym-explicit}
    E_\mathrm{asym} = \frac{U}{2}\left(\frac{\Delta}{\Delta+\Gamma}\right)^2.
\end{equation}

Notably, $E_\mathrm{asym}>0$ (since $U>0$). This means that Coulomb interaction pushes the energy of the odd state down with respect to the energies of states in the even sector. This tendency leads to the switch of the ground state parity from even to odd for sufficiently strong Coulomb repulsion \cite{yeyati2011}.
The energy separation between the even states is minimal at $\varphi = \pi$ and $\epsilon_g = 0$. In the vicinity of this point, even a weak interaction can render the odd state to be the ground state of the system [see Fig.~\ref{fig:energy}].
Combining Eqs.~\eqref{eq:EA} and \eqref{eq:asym-explicit} for $U\ll\Gamma + \Delta$ and $E_\mathrm{A}\ll\Delta$ we reproduce the known result for the boundary between the phases with even and odd ground states \cite{meng2009, novotny2015, novotny2019}.

Finally, we note that the asymmetry between the level-spacings can be probed in the tunneling spectroscopy of the junction \cite{pillet2010, chang2013} or in its microwave response, see Sec.~\ref{sec:asym} for the discussion of the latter approach. Measurement of $E_\mathrm{asym}$ can be used to experimentally assess the strength of the on-site Coulomb repulsion in the weak link.

\section{Low-energy theory}
\label{sec:low-energy}
If the ABS is located well below the gap, $E_{\rm A} \ll \Delta$, the dynamical properties of the junction at small frequencies, $\omega \ll \Delta$, can be described with a help of a low-energy theory. Here we present such a theory. The requirement $E_{\rm A}\ll \Delta$ is fulfilled if $\Gamma, |\epsilon_g| \ll \Delta$. It is also fulfilled for arbitrary $\Gamma/\Delta$ if $|\Gamma_L - \Gamma_R|, |\epsilon_g| \ll \Delta + \Gamma$ and at the same time $|\varphi - \pi| \ll 1$.

The fermion number parity is conserved within our model. Thus, the dynamics of the system can be studied separately for states with odd and even fermion parity. In the odd parity sector, states $|1_\uparrow\rangle$ and $|1_\downarrow\rangle$ are not coupled by the applied phase or gate voltage drives due to spin conservation. Thus, in the odd states the system adiabatically follows the change in $\epsilon_g$ and $\varphi$ induced by the drives as long as the drive frequency is small, $\omega\ll\Delta$.

The dynamics is more intricate in the even parity sector. If the frequency of the drives is comparable to $E[2]-E[0] = 2E_\mathrm{A}$, the transitions between states $|0\rangle$ and $|2\rangle$ have to be accounted for. This dynamics can be captured by a low-energy Hamiltonian. The latter can be obtained from the full Hamiltonian by applying a two-level adiabatic approximation.
In the particle-hole basis, the low-energy Hamiltonian is given by (see Appendix \ref{sec:app-low-en-ham} for the detailed derivation)
\begin{equation}
    \label{eq:low-en}
    H^{(\mathrm{le})}_\mathrm{even} = E_{\mathrm{even}}+ \frac{\Delta}{\Delta + \Gamma}  \begin{pmatrix}
    \tilde{\epsilon}(t)& \tilde{\gamma}[\varphi(t)]\\
    \tilde{\gamma}^\star[\varphi(t)] & -\tilde{\epsilon}(t)
    \end{pmatrix},
\end{equation}
where $\tilde{\epsilon}(t) =\tilde{\epsilon}_g(t) -\frac{\delta\Gamma}{2\Delta} e V(t)$
[with $V(t) = \phi_0 \partial_t \varphi(t)$ and $\delta\Gamma = \Gamma_L - \Gamma_R$]; $\tilde{\epsilon}_g$ and $\tilde{\gamma}[\varphi]$ are defined in Eq.~\eqref{eq:renorm}. The $c$-number contribution $E_{\mathrm{even}}$ is given by Eq.~\eqref{eq:even}. In the static case, the energy spectrum of Hamiltonian \eqref{eq:low-en} is given by Eq.~\eqref{eq:E02} with $E_{\rm A}$ of Eq.~\eqref{eq:EA}.

Hamiltonian \eqref{eq:low-en} has several notable features. First, particles and holes are coupled via the off-diagonal matrix element $\propto\tilde{\gamma}$. These pairing correlations originate due to the proximity effect arising from the superconducting leads.  Second, $\tilde{\epsilon}_g$ is attenuated by the factor of $\Delta/(\Delta + \Gamma) < 1$ which describes the probability of finding a quasiparticle at the dot (as opposed to the leads), see discussion after Eq.~\eqref{eq:low-en-abs}.
Finally, there is a peculiar correction $-\frac{\delta\Gamma}{2\Delta}e V(t)$ to the potential energy of the dot. It describes the average potential felt by the quasiparticle during its virtual excursions to the leads. Notably, this correction vanishes for $\Gamma_R=\Gamma_L$ since we assume $V_R(t)=-V_L(t) = V(t)/2$. Formally, such a correction to the Hamiltonian stems from Berry connection, $-\frac{i\hbar}{2}\langle p |\dot{p}\rangle+\frac{i\hbar}{2}\langle h|\dot{h}\rangle \approx -\frac{\delta\Gamma}{2\Delta}e V(t)$, where $|p\rangle$ and $|h\rangle$ are the particle and hole states respectively (see Appendix \ref{sec:app-low-en-ham} for details). 

It is convenient to perform a time-dependent gauge transformation that removes $V(t)$ from the diagonal components of Eq.~\eqref{eq:low-en}. This leads to $H^{(\mathrm{le})}_\mathrm{even} \rightarrow {H}^{\prime(\mathrm{le})}_\mathrm{even}$ with
\begin{equation}
    \label{eq:low-en2}
    H^{\prime(\mathrm{le})}_\mathrm{even} = E_{\mathrm{even}}+ \frac{\Delta}{\Delta + \Gamma}  \begin{pmatrix}
    \tilde{\epsilon}_g(t) & z[\varphi(t)]\\
    z^\star[\varphi(t)] & -\tilde{\epsilon}_g(t)
    \end{pmatrix},
\end{equation}
where
\begin{equation}
\label{eq:z}
z[\varphi] = \exp{\left(-i\frac{\varphi}{2}\frac{\delta\Gamma }{\Delta+\Gamma}\right)}\tilde{\gamma}[\varphi].
\end{equation}
One can check that in this gauge the low-energy charge and current operators can be obtained as the derivatives of the low-energy Hamiltonian: $\hat{Q}^{\mathrm{(le)}} = -e \partial_{\epsilon_g}H^{\prime\mathrm{(le)}}_{\mathrm{even}}$ and $\hat{I}^{\mathrm{(le)}} = \phi_0^{-1} \partial_\varphi H^{\prime\mathrm{(le)}}_\mathrm{even}$, respectively. This property is useful for studying the electromagnetic response of the system and therefore we always work with the gauge-transformed version of the low-energy Hamiltonian. Thus in what follows we omit the prime in $H^{\mathrm{\prime(le)}}_{\mathrm{even}}$.

We note that the limit $\Gamma\gg\Delta$ reduces the Hamiltonian \eqref{eq:low-en2} to that of a short junction with a reflection amplitude $r = \delta\Gamma/\Gamma$ \cite{zazunov2003}. Importantly, in this regime quasiparticles occupying the Andreev bound state predominantly stay within the leads (and not at the dot). Therefore, the drive applied to the gate cannot induce transitions within the even parity sector and all charging effects are suppressed.

Finally, we remind that Hamiltonian \eqref{eq:low-en} was derived under the assumption that capacitance between the dot and the gate is much larger than the capacitance between the dot and the leads. If the capacitances to the leads and to the gate are comparable, $\epsilon_g(t)$ [and thus $\tilde{\epsilon}_g$(t)] in Eq.~\eqref{eq:low-en2} starts to depend on voltages in the leads in addition to $V_g(t)$. We present this dependence in Appendix \ref{app:cap}.

\section{Linear response}
\label{sec:lin-res}
In this section, we study the linear electromagnetic response of the ABS to weak externally applied drives $\delta V_g(t)$ and $\delta\phi(t) = \phi_0\delta\varphi(t)$ [note that we use the flux variable $\phi$ to characterize the differential phase drive; $\phi_0 = \hbar/2e$ is the reduced flux quantum]. As follows from linearizing Hamiltonian \eqref{eq:model}, time-dependent perturbations describing these drives are given by $\delta H_{Q}(t) = \hat{Q} \delta V_g(t)$ and $\delta H_I (t)=  \hat{I}\delta\phi(t)$, respectively.
The linear response function $\chi[\omega, n] \equiv \chi[\omega, \epsilon_g, \varphi, n]$ depends on the state in which the system resides before the application of the perturbations. $n=0$ and $n=2$ correspond to states $|0\rangle$ and $|2\rangle$, respectively. $n=1$ corresponds to either $|1_\uparrow\rangle$ or $|1_\downarrow\rangle$ (the response function in our model does not depend on spin and we do not specify it in the definition of $\chi$). We define the response function as a matrix
\begin{gather}
    \begin{pmatrix}
    \delta Q(\omega)\\
    \delta I(\omega)
    \end{pmatrix}
    =
    \chi[\omega,n]
    \begin{pmatrix}
    \delta V_g(\omega)\\
    \delta\phi(\omega)
    \end{pmatrix},\:
    \chi =  \begin{pmatrix}
    \chi_{QQ}&\chi_{QI}\\
    \chi_{IQ}&\chi_{II}
    \end{pmatrix}.
\end{gather}
Here, $\delta Q$ and $\delta I$ are the deviations of the average charge and current from their stationary values. As usual, the Hermitian (anti-Hermitian) part of $\chi$ describes the non-dissipative (dissipative) response of the system. The response function matrix satisfies a general relation that guarantees that physical quantities are real, $\chi[\omega]=\chi^\star[-\omega]$.

Prior to computing $\chi$, we discuss symmetry properties of this matrix. From time-reversal and particle-hole symmetries we obtain
\begin{align}
    \chi[\omega,\varphi,\epsilon_g]=&M\chi[-\omega,-\varphi,\epsilon_g]M\label{eq:tr-symm},\\
     \chi[\omega,\varphi,\epsilon_g]=& \chi[\omega,-\varphi,-\epsilon_g]\label{eq:ph-symm},
\end{align}
respectively (for brevity we omitted the state argument $n$). Here matrix $M = \mathrm{diag}\{1,-1\}$. From inversion symmetry we get \footnote{Action of the inversion symmetry defined by Eq.~\eqref{eq:inv-symm} needs to be modified if the capacitances between the dot and the leads, $C_L$ and $C_R$, are comparable to the capacitance between the dot and the gate, $C_g$. In that case, under the action of inversion symmetry capacitances $C_L$ and $C_R$ should be exchanged similarly to $\Gamma_L$ and $\Gamma_R$.}
\begin{equation}
    \label{eq:inv-symm}
    \chi[\omega,\varphi,\epsilon_g,\Gamma_L,\Gamma_R]=M\chi[\omega,-\varphi,\epsilon_g,\Gamma_R,\Gamma_L]M.
\end{equation}
In this expression we introduced $\Gamma_L$ and $\Gamma_R$ as arguments. Note that in the right side of Eq.~\eqref{eq:inv-symm} these arguments are exchanged. The symmetry relations have a set of important consequences for the off-diagonal component of the response function $\chi_{IQ}$ (similar conclusions are true for $\chi_{QI}$). From Eq.~\eqref{eq:tr-symm} we see that $\mathrm{Re}\,\chi_{IQ}$ vanishes at time-reversal symmetric points $\varphi = 0,\pi$. As a consequence of particle-hole symmetry, $\mathrm{Re}\,\chi_{IQ}$ also vanishes at $\epsilon_g=0$. Finally, from Eqs.~\eqref{eq:tr-symm} and \eqref{eq:inv-symm} we see that $\mathrm{Im}\,\chi_{IQ}=0$ for an inversion-symmetric weak link, $\Gamma_R = \Gamma_L$, at any $\varphi$ and $\epsilon_g$.

{Now we proceed to the calculation of the response functions in the considered discrete states. The components of matrix $\chi$} can be expressed as (see Appendix \ref{app:general_lin_resp})
\begin{equation}
    \label{eq:response}
    \chi_{AB}[\omega,n] = \partial_{a}\partial_b E[n] + \delta\chi_{AB}[\omega, n].
\end{equation}
Here, indices $A$ and $B$ stand for $Q$ or $I$ while $a$ and $b$ are the respective drive variables, $V_g$ or $\phi$. The second term in Eq.~\eqref{eq:response} vanishes at zero frequency [see Eq.~\eqref{eq:delta-chi}]. The first term in Eq.~\eqref{eq:response} is, in contrast, non-zero at $\omega = 0$. It describes the frequency-independent adiabatic part of the response function (naturally, this contribution is purely non-dissipative). Its diagonal components are the inverse inductance of the junction, $1/L[n] = \partial^2_{\phi} E[n]=\phi_0^{-2}\partial^2_{\varphi} E[n]$, and the quantum capacitance, $C[n] = \partial^2_{V_g} E[n] = \partial_{V_g} Q[n]$, where $Q$ describes the average charge at the dot. The off-diagonal component, $\partial_{\phi}\partial_{V_g} E[n] = \partial_\phi Q[n] \equiv \phi_0^{-1}\partial_\varphi Q[n]$, describes how the charge at the dot changes with phase $\varphi$. It is also related to the change of the Josephson current with gate voltage, $\partial_{\phi}\partial_{V_g} E[n] = \partial_{V_g} I[n]$, such that Maxwell's relation holds, $\partial_\phi Q = \partial_{V_g} I$. Due to time-reversal symmetry, $\partial_{\phi}\partial_{V_g} E[n]$ vanishes at $\varphi = 0$ and $\varphi=\pi$. $\partial_{\phi}\partial_{V_g} E[n]$ also vanishes at $\epsilon_g=0$ due to the presence of the particle-hole symmetry.

The second term in Eq.~\eqref{eq:response} describes the dynamic part of the response function,
\begin{equation}
\label{eq:delta-chi}
\delta\chi_{AB}[\omega,n]=\chi_{AB}^\mathrm{K}[\omega,n]-\chi_{AB}^\mathrm{K}[0,n],
\end{equation}
where
$\chi_{AB}^\mathrm{K}[\omega]$ is given by the Kubo formula, $\chi_{AB}^\mathrm{K}[\omega] = -i \int_0^\infty dt e^{i\omega t} \langle [\hat{A}(t),\hat{B}]\rangle$. Here the average is computed over the unperturbed stationary state of the system which we assume to be either $|0\rangle$, $|1_\uparrow\rangle$, $|1_\downarrow\rangle$, or $|2\rangle$. The subtraction of the zero-frequency contribution is required to ensure that $\delta\chi_{AB}[0,n]=0$ and that the overall response function $\chi_{AB}$ is related to derivatives of energy at zero frequency (see Appendix~\ref{app:general_lin_resp} for detailed discussion). As follows directly from the Kubo formula, $\delta\chi_{AB}$ can be expressed as a sum over many-body states of the system,
\begin{equation}
    \label{eq:many-body-sum}
    \delta\chi_{AB}[\omega,n] = -\sideset{}{'}\sum_{k} \frac{\hbar\omega}{E_{kn}}\frac{A_{nk}B_{kn}}{E_{kn}-\hbar\omega-i0} + \mathrm{c.c.}(-\omega).
\end{equation}
Here, $k$ labels the many-body states, $E_{kn} = E[k] - E[n]$, prime designates that $k \neq n$, and $A_{nk}$ and $B_{kn}$ are the matrix elements of operators $\hat{A}$ and $\hat{B}$. Notice that the sum in Eq.~\eqref{eq:many-body-sum} runs over the states that belong to both discrete and continuous parts of the many-body spectrum. If the system is initially in an even state, $|0\rangle$ or $|2\rangle$, the sum involves one discrete state ($|2\rangle$ or $|0\rangle$, respectively) in addition to the states of continuum. If the system is in an odd state, $|1_\uparrow\rangle$ or $|1_\downarrow\rangle$, the sum includes only the states of the continuum; there are no matrix elements between $|1_\uparrow\rangle$ and $|1_\downarrow\rangle$ due to spin conservation.

We now describe how the response function can be computed in the limit of weak Coulomb interaction, $U \ll \Delta + \Gamma$, and small frequencies, $\hbar\omega \ll \Delta$ \footnote{More accurately, our derivation of the response functions is valid when $\hbar\omega \ll \Delta - E_\mathrm{A}$. For simplicity, in Section \ref{sec:lin-res} we focus on the limit $\Gamma \lesssim \Delta$ in which case $E_\mathrm{A} \lesssim \Delta$. Then it is enough to require $\hbar\omega\ll \Delta$, as is done in the main text.}. The weakness of interaction implies that the adiabatic part of the response function can be found using the perturbative expressions for the energies of the discrete states [see Eqs.~\eqref{eq:E02}, \eqref{eq:long-det}, \eqref{eq:even}, \eqref{eq:odd}].
The condition $\hbar\omega \ll \Delta$ implies that the terms in Eq.~\eqref{eq:many-body-sum} in which $k$ belongs to the many-body continuum are suppressed by a small parameter $\hbar\omega/\Delta$. This allows us to disregard $\delta\chi$ in comparison with the adiabatic part of the response function, $\partial_a\partial_b E$, if the system is initially in one of the odd states, $|1_\downarrow\rangle$ or $|1_\uparrow\rangle$.
The situation is different if the initial state is $|0\rangle$ or $|2\rangle$. Then, the sum in Eq.~\eqref{eq:many-body-sum} includes one discrete state in addition to the continuum.
The corresponding term may compete with the adiabatic part of the response function even at small frequencies $\hbar\omega \ll \Delta$, as long as $\hbar\omega \sim 2E_{\rm A}$.
Thus, we approximate the dynamic part of the response function in state $|0\rangle$ as
\begin{equation}
    \label{eq:Kubo}
    \delta\chi_{AB}[\omega,0] = -\frac{\hbar\omega}{2E_\mathrm{A}}\frac{A_{02}B_{20}}{2E_\mathrm{A}-\hbar\omega-i0} + \mathrm{c.c.}(-\omega).
\end{equation}
\begin{figure*}[t]
  \begin{center}
    \includegraphics[scale=1]{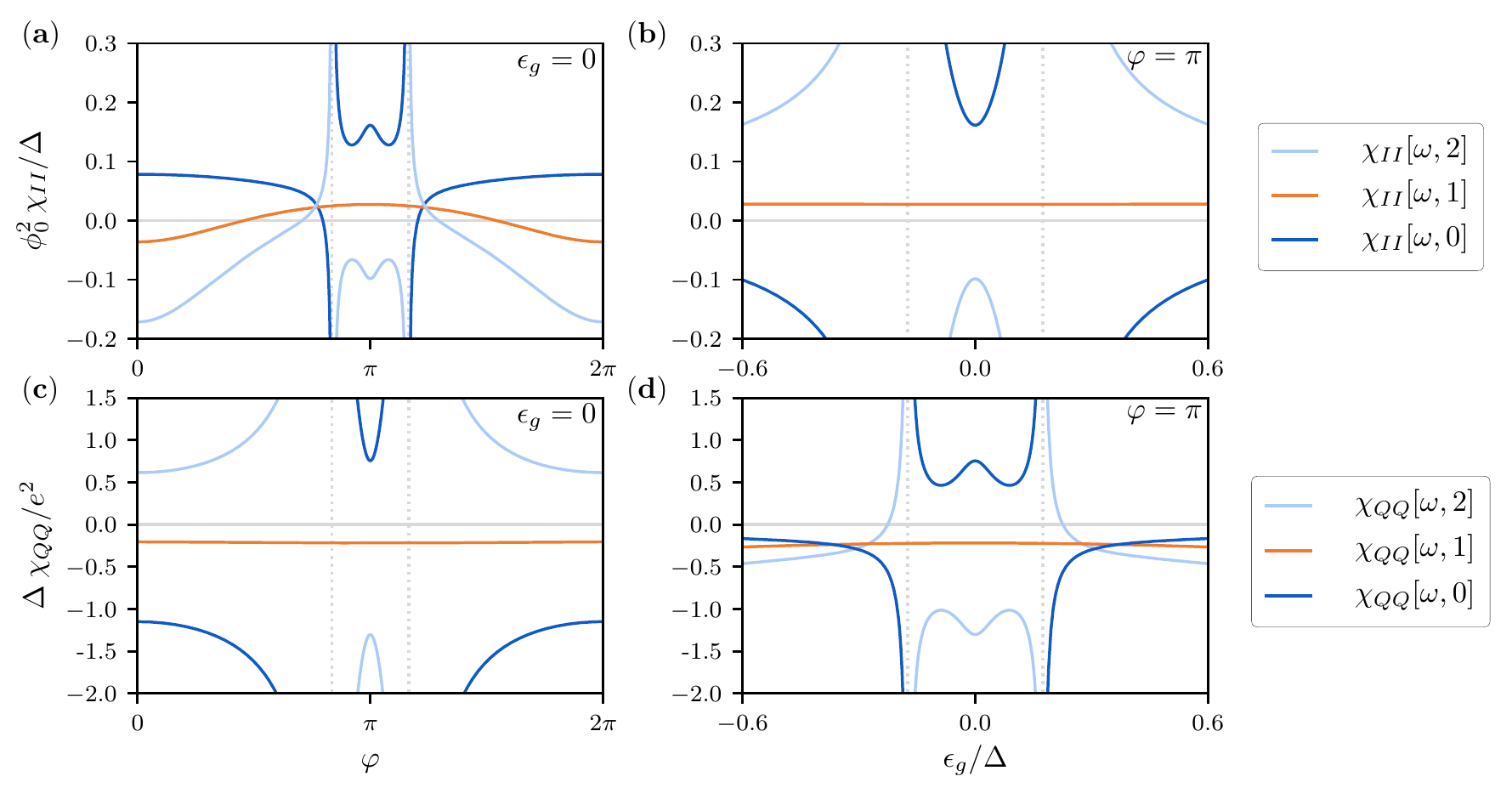}
    \caption{The response functions $\chi_{II}$ and $\chi_{QQ}$ of the ABS in states $|0\rangle$, $|1_\sigma\rangle$, $|2\rangle$ with different number of quasiparticles at the ABS. $\chi_{II}$ is plotted as a function of $\varphi$ in panel (a) [for $\epsilon_g = 0$] and as a function of $\epsilon_g$ in panel (b) [for $\varphi = \pi$]. $\chi_{QQ}$ is plotted as a function of $\varphi$ in panel (c) [for $\epsilon_g = 0$] and as a function of $\epsilon_g$ in panel (d) [for $\varphi = \pi$]. Plots are produced using Eqs.~\eqref{eq:response}, \eqref{eq:0=-2}, \eqref{eq:chi-QQ}, and \eqref{eq:chi-II}, for parameters $\Gamma_L = 0.3 \Delta$, $\Gamma_R = 0.35 \Delta$, $U = 0.35 \Delta$ (same as in Fig.~\ref{fig:energy}), and $\hbar\omega = 0.21 \Delta$. The response functions in states $|0\rangle$ and $|2\rangle$ diverge when the frequency is in resonance with the transition between $|0\rangle$ and $|2\rangle$, \textit{i.e.}, when $\hbar\omega = 2E_{\rm A}$ (vertical dashed lines in the plots). In the odd states and away from the resonances in the even states the response is adiabatic: $\chi_{II}\approx \partial_\phi^2 E$ describes the inverse inductance of the weak link and $\chi_{QQ} \approx \partial_{V_g}^2 E$ describes its quantum capacitance. The dissipative (imaginary) part of the response functions --- which is present at resonances only --- is not shown in the plot.}
    \label{fig:response}
  \end{center}
\end{figure*}
Notice that the response function has a resonant behavior at the transition frequency, $\hbar\omega = 2E_A$. Such a resonance corresponds to a process in which a drive photon is absorbed to change the occupation of the Andreev bound state. Away from the resonance, the low-frequency response is purely non-dissipative. The dynamic component of the response function in state $|2\rangle$ is approximately related to that in state $|0\rangle$ via
\begin{equation}
    \label{eq:0=-2}
    \delta\chi_{AB}[\omega,2] = -\delta\chi_{AB}[\omega,0],
\end{equation}
as follows directly from Eq.~\eqref{eq:many-body-sum} when neglecting terms with $k$ in the many-body continuum.

Next, recall that according to Eq.~\eqref{eq:Kubo} energy $E_\mathrm{A}$ should be comparable to $\hbar\omega$ for $\delta\chi_{AB}[\omega, 0]$ to produce an appreciable contribution to the response function $\chi_{AB}[\omega, 0]$. Since we assume $\hbar\omega \ll \Delta$, in such case $E_\mathrm{A}$ is also small and thus $\delta\chi_{AB}[\omega, 0]$ can be approximately computed with the help of the low-energy theory of Sec.~\ref{sec:low-energy}. Correspondingly, the charge and current operators in Eq.~\eqref{eq:Kubo} can be exchanged for their low-energy versions. Then, for the charge-charge component of the response function we find
\begin{equation}
    \delta\chi_{QQ}[\omega,0] = - e^2 \frac{\hbar^2\omega^2}{4E_\mathrm{A}^2-(\hbar\omega+i0)^2} \partial_{\epsilon_g}^2 E_{\mathrm{A}}.
    \label{eq:chi-QQ}
\end{equation}
Note that for $\hbar\omega \ll E_\mathrm{A}$, the response function scales as $\delta\chi_{QQ}[\omega,0] \propto \omega^2$. Similarly to $\delta\chi_{QQ}$, the components $\delta\chi_{QI}$, $\delta\chi_{IQ}$, and $\delta\chi_{II}$ can be found with the help of the low-energy Hamiltonian \eqref{eq:low-en2}. However, in general the resultant expressions are cumbersome and we relegate them to Appendix \ref{app:full-expr}. Here, we invoke an additional approximation to illustrate the qualitative features of the results. Namely, we disregard the phase factor in Eq.~\eqref{eq:z} since it gives only the subleading corrections of order $E_\mathrm{A}/\Delta \ll 1$ to the response functions. Neglecting such corrections, we find for $\delta\chi_{IQ}$
\begin{gather}
    \delta\chi_{IQ}[\omega, 0] = e\phi_0^{-1}\frac{1}{4 E_\mathrm{A}^2 - (\hbar\omega + i0)^2}\Bigg[\hbar^2\omega^2 \partial_{\epsilon_g}\partial_{\varphi} E_{\mathrm{A}}+\notag\\ + \left(\frac{\Delta}{\Delta + \Gamma}\right)^3\frac{i\hbar\omega}{E_{\mathrm{A}}}\Gamma\delta\Gamma\left(1+f \frac{U}{\Delta}\right)\left(1+g \frac{U}{\Delta}\right)^2\Bigg].
\label{eq:chi-IQ}
\end{gather}
Notice that there exists a well-defined limit $\delta\chi_{IQ}[\omega,n]/(-i\omega)|_{\omega \rightarrow 0} = C_p$ which describes the capacitive response of the polarization charge between the leads to the applied gate voltage. The capacitance $C_p$ vanishes for symmetric contacts, $\delta\Gamma = 0$. This is because in that case the system is symmetric under a combination of a time-reversal and inversion symmetries (see Eq.~\eqref{eq:tr-symm} and Eq.~\eqref{eq:inv-symm}, respectively). The response function $\delta \chi_{QI}$ can be obtained from Eq.~\eqref{eq:chi-IQ} by conjugating the expression in the bracket. Finally, for $\delta \chi_{II}$ within the adopted approximations we obtain
\begin{gather}
    \notag\delta\chi_{II}[\omega,0] =-\frac{\phi_0^{-2}}{E_\mathrm{A}} \frac{\hbar^2\omega^2}{4E_\mathrm{A}^2-(\hbar\omega+i0)^2}\frac{1}{|\tilde{\gamma}|^2}\Bigg[\tilde{\epsilon}_g^2 (\partial_\varphi E_\mathrm{A})^2 + \\   +\frac{1}{4}\left(\frac{\Delta}{\Delta+\Gamma}\right)^2 \Gamma^2\delta\Gamma^2\left(1+g \frac{U}{\Delta}\right)^4\Bigg].
    \label{eq:chi-II}
\end{gather}
Capacitance $\delta\chi_{II}[\omega,n]/(-i\omega)^2|_{\omega \rightarrow 0}$ describes the response of {the polarization charge to the voltage bias between the leads}. 
For $\Gamma\gg\Delta,|\delta\Gamma|,|\epsilon_g|, U$ equation \eqref{eq:chi-II} reduces to
\begin{equation}
\delta\chi_{II}[\omega,0] =-\frac{\phi_0^{-2}}{E_\mathrm{A}} \frac{\hbar^2\omega^2}{4E_\mathrm{A}^2-(\hbar\omega+i0)^2}\frac{\frac{1}{4}\Delta^2\delta\Gamma^2}{\Gamma^2 - 4\Gamma_R \Gamma_L \sin^2\frac{\varphi}{2}}.
\label{eq:chi-II-tunnel}
\end{equation}
This limit corresponds to the case of a short single-channel junction with high transparency. Accordingly, Eq.~\eqref{eq:chi-II-tunnel} reproduces the known result for the response function of the short junction near $\varphi = \pi$ \cite{kos2013}.

{We demonstrate the behavior of the low-frequency response functions $\chi_{QQ}[\omega, n]$ and $\chi_{II}[\omega, n]$ for a particular choice of model parameters in Fig.~\ref{fig:response} [plots for $\chi_{IQ}$ and $\chi_{QI}$ are presented in Appendix \ref{app:iq-and-qi}]. The parameters are chosen to demonstrate resonant behavior, $\hbar\omega = 2 E_{\rm A}$, at specific values of $\varphi$ and $\epsilon_g$. The response functions $\chi_{AB}[\omega, 0]$ and $\chi_{AB}[\omega, 2]$ diverge at the resonances and change sign across them. The response functions in the odd states $|1_\sigma\rangle$ are approximately adiabatic, $\chi_{AB}[\omega, 1] \approx \partial_a \partial_b E[1]$, since $\hbar\omega$ is small compared to $\Delta$. For weak Coulomb interaction, they are primarily determined by $E_\mathrm{cont}$, \textit{i.e.}, the contribution of the occupied continuum states [see Eqs.~\eqref{eq:unpert}, \eqref{eq:cont}, \eqref{eq:odd}]. Notice that at $\Gamma\sim \Delta$ the phase dependence of $\chi_{II}[\omega,1]$ --- which is mainly determined by the continuum contribution --- is comparable in magnitude to that of $\chi_{II}[\omega, 0]$ and $\chi_{II}[\omega, 2]$ [see Fig.~\ref{fig:response}(a)]. At the same time, $\chi_{II}[\omega,1]$ is almost independent of the gate voltage up to $|\epsilon_g| \sim \Delta$ [see Fig.~\ref{fig:response}(b)]. $\chi_{QQ}[\omega,1]$ weakly depends on both $\varphi$ and $|\epsilon_g|\lesssim \Delta$ even though $\Gamma \sim \Delta$ [see Fig.~\ref{fig:response}(c) and Fig.~\ref{fig:response}(d)].}

To conclude this section, we note that without the interaction, $U = 0$, the dynamic part of the response function can be calculated in our model exactly. The resultant expressions are cumbersome, so we present them in Appendix \ref{sec:app-exact}.

\section{Asymmetry of the response functions}
\label{sec:asym}
Results of Sections \ref{sec:spectrum}--\ref{sec:lin-res} indicate that the response functions are sensitive to the on-site Coulomb repulsion. Therefore, the measurement of these functions might be used to estimate the strength of the Coulomb interaction. A particularly convenient combination of the response functions that explicitly characterizes the magnitude of parameter $U$ is
\begin{equation}
\label{eq:chi-asym}
 \chi_{AB}^{\mathrm{asym}}[\omega] = \frac{\chi_{AB}[\omega,2]+\chi_{AB}[\omega,0]}{2} - \chi_{AB}[\omega,1]
\end{equation}
which we call the response asymmetry. This quantity is illustrative because it is non-zero only in the presence of Coulomb interaction.
The latter property can be easily seen in the limit $\omega \rightarrow 0$. Indeed, $\chi^\mathrm{asym}_{AB}[\omega \rightarrow 0]=\partial_a \partial_b E^\mathrm{asym}$ and $E^\mathrm{asym} = 0$ for $U = 0$ as was shown in Section \ref{sec:spectrum}. We demonstrate in Appendix \ref{sec:app-exact} that the response asymmetry also vanishes at $\omega \neq 0$ when $U = 0$.

\begin{figure}[t]
  \begin{center}
    \includegraphics[scale=1.0]{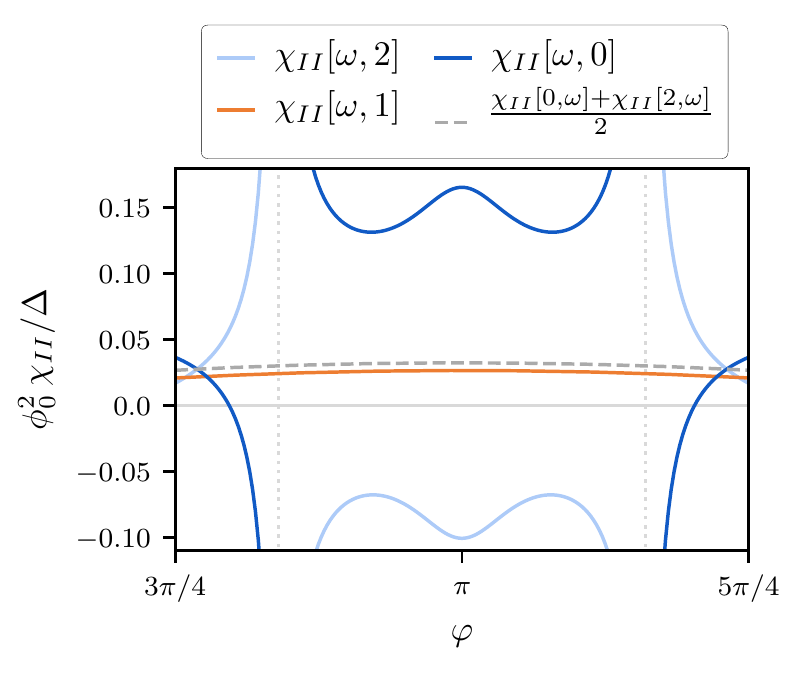}
    \caption{Asymmetry of the response functions $\chi_{II}[\omega, n]$. $\chi_{II}[\omega,n]$ is plotted as a function of phase in the vicinity of $\varphi = \pi$ for $\Gamma_L = 0.30 \Delta$, $\Gamma_R = 0.35 \Delta$, $U = 0.5 \Delta$, $\hbar\omega = 0.21 \Delta$.
    Vertical lines correspond to the resonances at $\hbar\omega = 2E_A$. Dashed line shows the half-sum between $\chi_{II}[\omega, 0]$ and $\chi_{II}[\omega, 2]$. Due to the presence of Coulomb interaction, the half-sum differs from $\chi_{II}[\omega, 1]$ by $\chi_{II}^\mathrm{asym}[\omega]$ [see Eq.~\eqref{eq:response-asymmetry}]. \label{fig:asym}}
  \end{center}
\end{figure}

We calculate $\chi_{AB}^\mathrm{asym}$ in the regime of perturbatively weak Coulomb repulsion and low frequency, $\hbar\omega \ll \Delta$.
There, the asymmetry reduces to that of the adiabatic components of the response functions. Indeed, in this limit the dynamic components of the response functions cancel in the combination $\chi_{AB}[\omega,0]+\chi_{AB}[\omega,2] $, as can be seen from Eq.~\eqref{eq:0=-2}. At the same time,  $\delta\chi_{AB}[\omega, 1]$ is small compared to the adiabatic part of the response function and can be disregarded [see the discussion after Eq.~$\eqref{eq:many-body-sum}$]. Thus,  $\chi^\mathrm{asym}_{AB}$ can be directly found from the asymmetry of the energies, see Eq.~\eqref{eq:en-asymmetry}. In this way, we obtain the approximate relation
\begin{equation}
\label{eq:response-asymmetry}
    \chi_{AB}^{\mathrm{asym}}[\omega] = \frac{U}{2}\partial_a \partial_b \alpha^2,
\end{equation}
where $\alpha$ is given by Eq.~\eqref{eq:alpha}. Expression \eqref{eq:response-asymmetry} can be simplified in the limit of weak coupling between the level and the leads, $\Gamma,|\epsilon_g| \ll \Delta$. To the lowest non-vanishing order we obtain
\begin{gather}
    \chi_{QQ}^{\mathrm{asym}} = - e^2\frac{3U}{\Delta}\frac{\Gamma}{\Delta^2},\quad \chi_{II}^{\mathrm{asym}}  = -\phi_0^{-2}\frac{2U}{\Delta}\frac{\Gamma_R \Gamma_L \cos\varphi}{\Delta},\notag\\
    \chi_{IQ}^{\mathrm{asym}} = \chi_{QI}^{\mathrm{asym}} = e \phi_0^{-1}\frac{8U}{\Delta} \frac{\Gamma_R\Gamma_L \epsilon_g \sin \varphi}{\Delta^3}.
\end{gather}
For stronger coupling, $\Gamma\lesssim\Delta$, we demonstrate the asymmetry of inductive response functions $\chi_{II}[\omega, n]$ in Fig.~\ref{fig:asym}. The asymmetry of inductive responses was recently measured in our experiment \cite{fatemi2021}, pointing to the importance of Coulomb interaction for the microwave properties of nanowire weak links.

\section{ABS in circuit QED}
\label{sec:cqed}
\begin{figure*}[t]
  \begin{center}
    \includegraphics[scale=0.833]{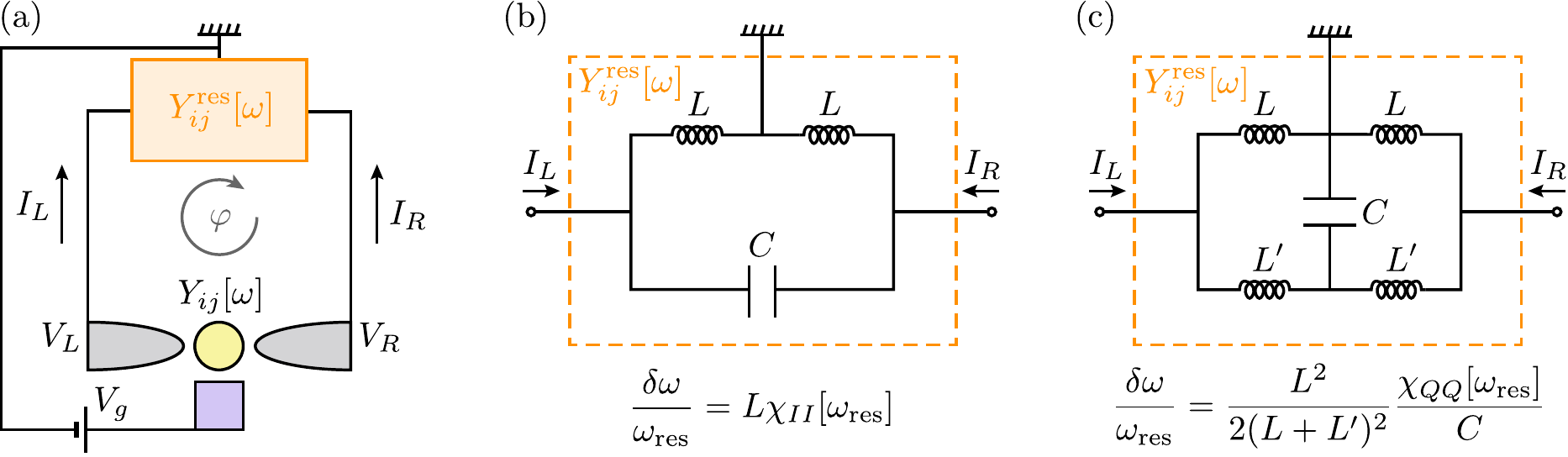}
    \caption{(a) A finite-length weak link with a state-dependent admittance $Y_{ij}[\omega]$ is galvanically connected to a microwave resonator with admittance $Y_{ij}^\mathrm{res}[\omega]$ (indices $i,j\in\{L,R\}$ denote the left or the right lead, respectively). An external flux tunes the phase difference $\varphi$ across the weak link. The gate voltage $V_g$ also tunes the energy of the fermionic level in the weak link. The presence of the weak link shifts the frequency of the mode of the resonator, $\omega_\mathrm{res} \rightarrow \omega_\mathrm{res} + \delta \omega$. (b) Effective circuit representing a microwave resonator which can be used to measure the response function $\chi_{II}$ of the weak link. Due to the symmetry of the resonator {with respect to the ground}, its mode has opposite voltages on the left and at the right node. Then, using Eq.~\eqref{eq:dw}, we find $\delta\omega/\omega_\mathrm{res} =  L\chi_{II}[\omega_\mathrm{res}]$ as long as $\chi_{II}\ll L^{-1}$. (c) Effective circuit representing a microwave resonator which can be used to probe the response function $\chi_{QQ}$ of the weak link. The mode of the resonator has the same voltages at the left and at the right nodes which, according to Eq.~\eqref{eq:dw}, leads to $\delta\omega/\omega_\mathrm{res} = L^2\chi_{QQ}[\omega_\mathrm{res}]/[2 (L+L^\prime)^2 C]$ for $\chi_{QQ}\ll C$.}
    \label{fig:resonators}
  \end{center}
\end{figure*}

Above we demonstrated that the microwave response of a finite-length weak link is characterized by four response functions, $\chi_{QQ}$, $\chi_{QI}$, $\chi_{IQ}$, and $\chi_{II}$. Experimentally, these response functions can be studied using the toolbox of circuit quantum electrodynamics (cQED). In cQED, the weak link coupled to a microwave resonator shifts the frequency of the latter. The magnitude of this dispersive shift may be related to a certain combination of the response functions, specific for a particular resonator design. In this section, we elucidate this relation and demonstrate how different response functions can be measured by appropriately tailoring the geometry of the resonator.

We assume that the weak link hosting ABS is attached to a microwave resonator at two sites, $L$ and $R$ [see Fig.~\ref{fig:resonators} (a)]. External flux threads the loop between the junction and the resonator thus controlling the phase bias $\varphi$ across the weak link \footnote{The phase $\varphi$ drops at the weak link provided that inductance of the latter is much smaller than that of the resonator. Throughout Section \ref{sec:cqed} we assume that this condition is satisfied.}.
The resonator is modelled as a black box with a given matrix admittance, $Y^\mathrm{res}_{ij}[\omega]$, where $i,j \in \left\{L,R\right\}$. The admittance determines the relation between the currents flowing in the resonator and the voltages at nodes $L$ and $R$,
\begin{equation}
    I_i(\omega) = \sum_{j=L,R} Y^\mathrm{res}_{ij}[\omega] V_j(\omega).
\end{equation}
{We assume that the resonator is grounded (see Fig.~\ref{fig:resonators}) and that $V_L$ and $V_R$ are evaluated relative to the ground. Due to the presence of the ground, the currents explicitly depend on both $V_L$ and $V_R$ (\textit{i.e.}, not only on the voltage difference $V_L - V_R$).}
We assume that the photon loss in the resonator can be neglected such that $Y^\mathrm{res}_{ij}$ is an anti-Hermitian matrix. {The frequency of the modes of the unloaded resonator (\textit{i.e.}, in the absence of the weak link)} can be found as solutions of the characteristic equation
\begin{equation}
\label{eq:X}
    \mathrm{det}\,Y^\mathrm{res}_{ij}[\omega] = 0.
\end{equation}
Let $\omega_\mathrm{res}$ be the frequency of a given mode of the resonator determined by Eq.~\eqref{eq:X}. Then, the structure of the mode can be found by solving $\sum_{j=L,R}Y^\mathrm{res}_{ij}[\omega_\mathrm{res}] V_{\mathrm{res},j} = 0$.

When the weak link is present in the circuit, the characteristic equation changes to
\begin{equation}
\label{eq:Xx}
\mathrm{det}\left(Y^\mathrm{res}_{ij}[\omega] + Y_{ij}[\omega]\right) = 0.
\end{equation}
{Here $Y_{ij}[\omega]$ is the admittance of the resonant Andreev level.} The admittance matrix is related to the response functions computed in Section \ref{sec:lin-res} [see Eq.~\eqref{eq:response}] via 
\begin{subequations}
\label{eq:admit-1}
    \begin{align}
        Y_{LL}&= \frac{\chi_{II}}{-i\omega}+\frac{i\omega}{4}\chi_{QQ} + \frac{1}{2}\left(\chi_{IQ}-\chi_{QI}\right)\\
        Y_{LR}&= \frac{\chi_{II}}{i\omega}+\frac{i\omega}{4}\chi_{QQ} - \frac{1}{2}\left(\chi_{IQ}+\chi_{QI}\right),\\
        Y_{RL}&= \frac{\chi_{II}}{i\omega}+\frac{i\omega}{4}\chi_{QQ} + \frac{1}{2}\left(\chi_{IQ}+\chi_{QI}\right),\\
        Y_{RR}&= \frac{\chi_{II}}{-i\omega}+\frac{i\omega}{4}\chi_{QQ} - \frac{1}{2}\left(\chi_{IQ}-\chi_{QI}\right),        
    \end{align}
\end{subequations}
{(we suppressed the frequency arguments for brevity). We assume that the presence of the weak link does not affect the structure of the modes of the resonator.} This assumption is justified if the admittance of the load is small enough. In that case, Eq.~\eqref{eq:Xx} can be solved by taking $Y_{ij}$ into account perturbatively. We find that the frequency of the mode of the resonator in the presence of the load is given by $\omega_\mathrm{res} + \delta \omega$, where
\begin{equation}
\label{eq:dw}
\delta\omega =  - \frac{ \sum_{i,j\in L,R} V_{\mathrm{res},i}^\star Y_{ij}[\omega_\mathrm{res}] V_{\mathrm{res},j}}{\sum_{i,j\in L,R}  V_{\mathrm{res},i}^\star (Y^\mathrm{res}_{ij})^\prime[\omega_\mathrm{res}]V_{\mathrm{res},j}}
\end{equation}
and $(Y^\mathrm{res}_{ij})^\prime[\omega_\mathrm{res}] = dY_{ij}^\mathrm{res}/d\omega|_{\omega=\omega_\mathrm{res}}$. Equation \eqref{eq:dw} is a generalization of the relation between the frequency shift and the admittance \cite{devoret1989} to the multi-terminal case.
{The numerator of Eq.~\eqref{eq:dw} depends on $V_\mathrm{res}$, \textit{i.e.}, on the structure of the mode of the resonator. This opens a prospect of extracting particular components of the response function of the weak link by choosing a suitable geometry of the resonator.} In Figures \ref{fig:resonators} (b) and (c) we demonstrate lumped element circuits of resonators that can be used to isolately measure $\chi_{II}$ and $\chi_{QQ}$. We note the components $\chi_{QI}$ and $\chi_{IQ}$ cannot be measured separately from $\chi_{QQ}$ or $\chi_{II}$ irrespective of the geometry of the resonator (as can be verified directly from Eq.~\eqref{eq:dw}).

Finally, we mention that Eq.~\eqref{eq:dw} was derived under the assumption that the capacitance between the dot in the weak link and the gate is much larger than that between the dot and the leads. We derive a similar expression for arbitrary ratio of capacitances in Appendix~\ref{app:cap}, see Eq.~\eqref{eq:admit}.

\section{Conclusion}
Our work elucidates the problem of computing the microwave response of a finite-length weak link hosting a single ABS. Within a minimal Hamiltonian model, we calculated the corresponding linear response functions and found their evolution with the number of quasiparticles occupying ABS.
The resulting linear response functions can be used to analyze the state-dependent dispersive shifts in cQED experiments with the weak link coupled to a microwave resonator. Our minimal model captures the essential features differentiating a finite-length weak link from a point contact: (i) for the former both the ABS and the delocalized states contribute to the inductive response, and (ii) a finite-length link may accommodate electric charge making Coulomb interaction important. Our theory combined with the recent experimental results \cite{fatemi2021} highlight that quantum dot models provide an insightful perspective on microwave experiments with nanowire weak links. Below, we discuss the salient points of our work.

\textbf{Low-energy Hamiltonian.} The energy of an ABS formed in a finite-length weak link lies within the superconducting gap and does not reach $\Delta$ at any phase bias.
To describe the ABS energy spectrum and dynamics, we derived a
$2\times 2$ low-energy Hamiltonian which takes into account weak charging effects and delocalization of the quasiparticle into the leads [see Eq.~\eqref{eq:low-en2}]. This Hamiltonian yields an approximate expression for the ABS energy, see Eq.~(\ref{eq:func-form}),
where the dependence of $\delta$ and $\tau$ on the model parameters is presented in Eqs.~\eqref{eq:EA} and (\ref{eq:renorm}).
Equation \eqref{eq:func-form} becomes exact if
the tunneling rate is either large, $\Gamma\gg\Delta$, or small, $\Gamma\ll\Delta$ \cite{marcus2020, kou2020}.
It remains valid in the intermediate regime, $\Gamma\sim\Delta$, in the domain of $\varphi$ for which $E_\mathrm{A}(\varphi)\ll\Delta$, and provides a reasonable extrapolation between the solvable limits at $E_\mathrm{A}(\varphi) \sim \Delta$.

{The low-energy Hamiltonian describes the dynamics of the ABS in the even fermion parity sector. It provides a lumped element model of the weak link that can be used to analyze a variety of microwave experiments. Here, we applied the Hamiltonian to calculate the linear response functions of the weak link. The low-energy Hamiltonian may also be used in more complicated situations where the quantum fluctuations of phase across the weak link are appreciable~\cite{kou2020, marcus2020}.}

\textbf{Inductance of the continuum part of the weak-link spectrum.} As well-known in theory \cite{zazunov2003,feigelman1999,kos2013} and demonstrated in experiments \cite{janvier2015, metzger2021}, the dynamic response of a single-channel point contact is fully determined by the properties of the two-level ABS system hosted by the weak link. This is however not the case for the finite-length weak links. In particular, to compute the inductance of the weak link it is not enough to account for the contribution of the ABS $\propto(\partial^2_\varphi E_\mathrm{A})^{-1}$. This is because the continuum of states outside of the superconducting gap also contributes to the energy [see Eq.~\eqref{eq:cont}] and hence to the inductance. We show that the continuum contribution becomes comparable to that of the ABS when the coupling between the dot and the leads is strong, $\Gamma\sim\Delta$ [cf.~Eq.~\eqref{eq:low-en-abs} and Eq.~\eqref{eq:cont-weak-coupling}]. The effect of the continuum is especially prominent in the odd states, where it fully determines the inductance. Interestingly, the contribution of the continuum to the energy is minimal at phase bias $\varphi = \pi$ [see Eq.~\eqref{eq:cont-weak-coupling}]. A $\pi$-junction is thus realized whenever the ABS traps a single quasiparticle \footnote{The $\pi$-junction behavior for odd occupancy of the ABS should not be confused with that in the doubly occupied even state. In the latter case, the flipped energy-phase relation stems from the bound state contribution and is present even in a short single-channel junction. $\pi$-junction behavior in the odd state, in contrast, stems from the continuum states and is thus manifestly a finite-length effect.}. These results indicate that taking the continuum contribution into account is necessary to accurately describe microwave experiments with ABSs~\cite{fatemi2021}.

\textbf{Electrodynamic response functions.} A finite-length weak link may accommodate charge. Both this charge and the current through the weak link respond to gate voltage and phase bias. Thus, the microwave response of an ABS has a  multi-terminal character; the linear response functions form a $2\times 2$ matrix, $\chi$. At small frequencies the response functions are adiabatic, \textit{i.e.}, they can be found as the derivatives of the energies of the many-body states with respect to appropriate parameters [see Eq.~\eqref{eq:response}]. Adiabatic response functions characterize the quasi-static properties of the weak link, such as the inverse inductance and quantum capacitance. At finite frequency, a dynamic contribution to the response functions appears [see Eq.~\eqref{eq:delta-chi}]; it is the most prominent in the even sector where $\chi$ may have a resonant behavior [see Fig.~\ref{fig:response}]. Our theory smoothly interpolates between the adiabatic and resonant limits. The matrix $\chi$ can be accessed in cQED architecture by coupling the weak link to a microwave resonator and measuring the dispersive shift of the latter [see Eq.~\eqref{eq:admit-1} and \eqref{eq:dw}]. Recent experiments used this technique to study the inductive response of weak links \cite{metzger2021, fatemi2021}. The investigation of the capacitive response may be an interesting direction for future experimental works. Such a study can be carried out in a setup similar to that used in \cite{metzger2021, fatemi2021}, with appropriately modified resonator geometry [see Fig.~\ref{fig:resonators}(c)]. We note that interpolation between the adiabatic and resonant limits for the dispersive shift was also pointed out in a recent work \cite{park2020}.

\textbf{Coulomb interaction.} Accumulation of charge in a finite-length weak link makes the effects of Coulomb interaction important for the ABS physics. Surprisingly, most of the microwave experiments with nanowire weak links completely ignore the role of interaction in interpreting the data. Our recent experimental results \cite{fatemi2021} suggest that the on-site interaction may in fact be important to adequately describe the state-dependent response functions of the system. The theory presented here provides a guidance for assessing the strength of the interaction by comparing to each other the microwave responses measured at different occupancy of the ABS [see Eq.~\eqref{eq:chi-asym} and Eq.~\eqref{eq:response-asymmetry}].

Our quantum-dot inspired description of a finite-length weak link can be extended in multiple ways. In particular, it is possible to include the Zeeman effect, account for more levels in the weak link, and for spin-orbit interaction.
Looking forward, it would be interesting to compute and analyze, in the same framework, the microwave response of a finite-length weak link connecting topological superconductors \cite{fu2009, vayrynen2015}. It would also be interesting to evaluate the microwave response of a weak link containing a quantum dot in the regime of strong Coulomb interaction \cite{glazman2015, meng2009, novotny2015, novotny2019, yeyati2011, pillet2013, lee2012, chang2013}.

\acknowledgments{
We acknowledge very useful discussions with Steven Girvin, Manuel Houzet, Max Hays, Nick Frattini, Spencer Diamond, and Tom Connolly. We also acknowledge useful comments by Gianluigi Catelani. 
This work is supported by the ARO under Grant Number W911NF-18-1-0212.}
\bibliography{main}

\clearpage
\appendix

\widetext
\section{Spectral properties of the ABS in the absence of Coulomb interaction}
\label{app:non-int}

\subsection{Green's functions in the non-interacting case\label{app:gf}}
In this Appendix we derive the expression for the Green's functions of the system in the absence of interaction. The central result is the Green's function of the dot, $G_{dd}(\varepsilon)$ [see Eq.~\eqref{eq:G-inv} of the main text]. This Green's function can be conveniently used to find the energy of the Andreev bound state $E_\mathrm{A,0}$ [see Eq.~\eqref{eq:abs_unp_energy}], as well as the continuum contribution to the energy of the system, see Eq.~\eqref{eq:cont} and Appendix~\ref{app:mb_spectrum}. 
We start by rewriting the Hamiltonian \eqref{eq:model} in the particle-hole representation for $U = 0$. Up to an irrelevant $c$-number we obtain
\begin{equation}
\label{eq:ham-appendix}
    H = \sum_{i=L,R} \left(H_i+H_{T,i}\right) + H_d,
\end{equation}
where
\begin{equation}\label{eq:ham-appendix2}
    H_{i}	=\int d\mathbf{r}\,\Psi_{i}^{\dagger}(\mathbf{r})\left[\hat{\xi} \tau_z + \Delta \tau_x\right]
   \Psi_{i}(\mathbf{r}),\quad\quad
    H_{d} = D^{\dagger}\epsilon_g \tau_z D,\quad\quad H_{T,i}=t_{i}\left\{ D^{\dagger}\tau_{z}e^{\frac{i}{2}\tau_z\varphi_i}\Psi_{i}(0)+\mathrm{h.c.}\right\},
\end{equation}
$\tau_{x,y,z}$ are Pauli matrices in the Nambu space, $\Psi_i(\mathbf{r}) = (\psi_{i,\uparrow}(\mathbf{r}),\, \psi_{i,\downarrow}^{\dagger}(\mathbf{r}))^T$, and $D= (d_{\uparrow},\, d_{\downarrow}^{\dagger})^T$.
Next, we introduce the retarded and advanced  Green's functions $G^{R/A}$. These Green's functions have multiple components, of which the important ones are:
\begin{gather}
    G_{dd,\mu\nu}^{R/A}(t) = \mp i\theta(\pm t)\langle\{D_\mu(t), D_\nu^\dagger(0)\}\rangle,\quad\quad G_{ij,\mu\nu}^{R/A} = \mp i\theta(\pm t)\langle\{\Psi_{i,\mu}(0,t), \Psi_{j,\nu}^\dagger(0,0)\}\rangle,\\
    G_{id,\mu\nu}^{R/A}(t) = \mp i\theta(\pm t)\langle\{\Psi_{i,\mu}(0,t), D_\nu^\dagger(0)\}\rangle,\quad\quad G_{di,\mu}^{R/A}(t) = \mp i\theta(\pm t)\langle\{D_{
    \mu}(t), \Psi_{i,\nu}^\dagger(0,0)\}\rangle,
\end{gather}
where curly brackets denote the anticommutator. In the subsequent calculations we will only need the Green's functions at the position of the junction; hence the lead operators are all evaluated at $\mathbf{r} = 0$ in the above definitions. Using the Heisenberg equations of motion for the operators $D(t)$ and $\Psi_i(t)$, we may obtain a system of two coupled equations for $G_{dd}^{R/A}$ and $G_{id}^{R/A}$. In the energy domain, the system reads:
\begin{equation}
\label{eq:wf-eqns}
    (\varepsilon - \epsilon_g \tau_z \pm i0)G^{R/A}_{dd}(\varepsilon) = 1 + \sum_{i = L,R} T_i^\dagger G_{id}^{R/A}(\varepsilon),\quad  G_{id}^{R/A}(\varepsilon) = \frac{1}{V}\sum_k \frac{1}{\varepsilon - \xi_k\tau_z + \Delta\tau_x \pm i0}T_i G_{dd}^{R/A}(\varepsilon),
\end{equation}
where $T_i = t_i \tau_z e^{-\frac{i}{2}\tau_z \varphi_i}$. Substituting the second equation into the first and computing the sum over the momenta we obtain Eq.~\eqref{eq:G-inv} of the main text for $G_{dd}^{R/A}$ [in Eq.~\eqref{eq:G-inv}, $\varepsilon$ should be changed to $\varepsilon\pm i0$ for retarded and advanced Green's function, respectively].

For future reference, we also obtain the remaining components of the Green's functions. From Eq.~\eqref{eq:wf-eqns} we find
\begin{equation}\label{eq:id_app}
    G_{id}^{R/A}(\varepsilon) = g^{R/A}_\varepsilon T_i G^{R/A}_{dd}(\varepsilon),\quad g^{R/A}_\varepsilon = \frac{-\pi\nu}{\sqrt{\Delta^2 - (\varepsilon \pm i0)^2}}
    \begin{pmatrix}
    \varepsilon & \Delta \\
    \Delta & \varepsilon
    \end{pmatrix},
\end{equation}
where $\nu$ is the normal-state density of states in the leads (per spin projection). The components $G_{di}^{R/A}$ and $G_{ij}^{R/A}$ can be found analogously to $G_{dd}^{R/A}$ and $G_{id}^{R/A}$. We obtain
\begin{equation}
    \label{eq:exact-greens}
    G_{di}^{R/A}(\varepsilon) =   G^{R/A}_{dd}(\varepsilon) T_i^\dagger g^{R/A}_\varepsilon,\quad   G_{ij}^{R/A}(\varepsilon) = g^{R/A}_\varepsilon \delta_{ij} + g^{R/A}_\varepsilon T_i G^{R/A}_{dd}(\varepsilon) T_j^\dagger g^{R/A}_\epsilon.
\end{equation}

\subsection{Many-body energy spectrum \label{app:mb_spectrum}}
In this Appendix we describe the structure of the discrete many-body states $|0_0\rangle$, $|1_{0,\uparrow/\downarrow}\rangle$, and $|2_{0}\rangle$ in the absence of Coulomb interaction, and derive Eqs.~\eqref{eq:unpert}, \eqref{eq:cont} for the energies of these states. 

At $U = 0$, the many-body Hamiltonian [see Eqs.~\eqref{eq:ham-appendix}, \eqref{eq:ham-appendix2}] can be decomposed into the quasiparticle creation and annihilation operators as
\begin{equation}
\label{eq:non-int-ham}
H = \sum_{|\epsilon|>\Delta} \epsilon \gamma_\epsilon^\dagger \gamma_\epsilon + E_\mathrm{A,0} \gamma_{E_\mathrm{A,0}}^\dagger\gamma_{E_\mathrm{A,0}} - E_\mathrm{A,0} \gamma_{-E_\mathrm{A,0}}^\dagger\gamma_{-E_\mathrm{A,0}}.
\end{equation}
Here, we work in the ``semiconductor'' picture of superconductivity, in which there are states with both positive and negative energies. The first term corresponds to the states of the continuum. The second and the third terms describe the ABS; $E_{\rm A,0}$ is the ABS energy which may be found as a solution of $\det G^{-1}_{dd}(\varepsilon) = 0$ in the interval $\varepsilon \in [0, \Delta)$. The ground state of Hamiltonian~\eqref{eq:non-int-ham} corresponds to all single-particle states with negative energy being occupied,
\begin{equation}
    \label{eq:non-int-gs}
    |0_0\rangle = \gamma^\dagger_{-E_\mathrm{A,0}} |\mathcal{O}\rangle,\quad\quad\quad\quad\quad|\mathcal{O}\rangle = \prod_{\epsilon<-\Delta} \gamma_\epsilon^\dagger|\Omega\rangle,
\end{equation}
where $|\Omega\rangle$ is the vacuum state. Other discrete many-body states are
\begin{equation}
    \label{eq:app-states}
    |2_0\rangle = \gamma^\dagger_{E_\mathrm{A,0}}\gamma_{-E_\mathrm{A,0}}|0_0\rangle \quad\quad\quad|1_{0,\uparrow}\rangle = \gamma^\dagger_{E_\mathrm{A,0}} |0_0\rangle,\quad\quad\quad |1_{0,\downarrow}\rangle = \gamma_{-E_\mathrm{A,0}} |0_0\rangle.
\end{equation}
The number $n$ in the label of the state $|n_0\rangle$ characterizes the number of single-particle excitations above the ground state.

We can find energies of the discrete states with the help of Eqs.~\eqref{eq:non-int-ham}, \eqref{eq:non-int-gs}, and \eqref{eq:app-states},
\begin{equation}
    E_0[0/2] = E_{\rm cont} \mp E_{\rm A,0},\quad\quad E_0[1] = E_{\rm cont}.
\end{equation}
Here $E_{\rm cont} = \sum_{\epsilon < - \Delta} \epsilon$ corresponds to the total energy of the filled states of the continuum. This continuum contribution can be conveniently represented in terms of the Green's functions $G^{R/A}$ of the system, 
\begin{equation}\label{eq:cont_deriv}
    E_{\rm cont} = - \int_{-\infty}^{-\Delta}\frac{\varepsilon^\prime d\varepsilon^\prime}{2\pi i} \sum_\epsilon [-2\pi i\,\delta(\varepsilon^\prime - \epsilon)] = -\int_{-\infty}^{-\Delta}\frac{\varepsilon^\prime d\varepsilon^\prime}{2\pi i} \frac{\partial}{\partial \varepsilon^\prime}\ln \det \left[G^A (\varepsilon^\prime) \left[G^R(\varepsilon^\prime)\right]^{-1}\right].
\end{equation}
As can be shown,
\begin{equation}\label{eq:repr_app}
    \det G^{R/A}(\varepsilon^\prime) = \det G^{R/A}_{dd}(\varepsilon^\prime) \cdot \det \mathfrak{g}^{R/A}_{R, 0}(\varepsilon^\prime) \cdot \det \mathfrak{g}^{R/A}_{L, 0}(\varepsilon^\prime),
\end{equation}
where $\mathfrak{g}^{R/A}_{i, 0}$ are the Green's functions of the lead $i$ in the absence of tunneling to the dot [it is given by $\mathfrak{g}^{R/A}_{i, 0}(\varepsilon^\prime) = (\varepsilon^\prime\pm i0 - [\hat{\xi}\tau_z + \Delta \tau_x])^{-1}$]. Substituting representation~\eqref{eq:repr_app} in Eq.~\eqref{eq:cont_deriv} we obtain
\begin{equation}
    E_{\rm cont} = -\sum_{i = L, R} \sum_k \sqrt{\Delta^2 + \xi_k^2} \,-\,\int_{-\infty}^{-\Delta}\frac{\varepsilon^\prime d\varepsilon^\prime}{2\pi i} \frac{\partial}{\partial \varepsilon^\prime}\ln \det \left[G_{dd}^A (\varepsilon^\prime) \left[G_{dd}^R(\varepsilon^\prime)\right]^{-1}\right],
\end{equation}
While the first term is thermodynamically large, it is also independent of the phase and gate-voltage biases $\varphi$, $\epsilon_g$. Therefore, it has no effect on the dynamics of the ABS. By omitting the irrelevant first term and performing integration by parts on the second term, we obtain Eq.~\eqref{eq:cont}.

\section{Corrections to energies due to the interaction}
\label{app:int}
In this Appendix we find the first-order corrections to the energies of the discrete states $|0_0\rangle$, $|1_{0,\sigma}\rangle$, and $|2_0\rangle$ due to a finite strength of Coulomb interaction $U \ll \Delta + \Gamma$.

According to the first-order perturbation theory, we need to compute the matrix elements of the interaction Hamiltonian
\begin{equation}\label{eq:Hint_app}
H_\mathrm{int} = U\bigl(d_{\uparrow}^{\dagger}d_{\uparrow}-\frac{1}{2}\bigr)\bigl(d_{\downarrow}^{\dagger}d_{\downarrow}-\frac{1}{2}\bigr)
\end{equation}
between the unperturbed discrete states. To do that, it is convenient to expand $H_\mathrm{int}$ into the quasiparticle operators $\gamma_\epsilon$ [see Eq.~\eqref{eq:ham-appendix}]. The expansion may be performed with the help of the eigenstate decompositions of the creation and annihilation operators for an electron at the dot:
\begin{equation}
\label{eq:ph-operators-1}
d_\uparrow =p_{E_\mathrm{A,0}}\gamma_{E_\mathrm{A,0}}+p_{-E_\mathrm{A,0}}\gamma_{-E_\mathrm{A,0}}+\sum_{|\epsilon|>\Delta}p_\epsilon \gamma_\epsilon, \quad\quad\quad d_\downarrow^\dagger = h_{E_\mathrm{A,0}}\gamma_{E_\mathrm{A,0}}+h_{-E_\mathrm{A,0}}\gamma_{-E_\mathrm{A,0}}+\sum_{|\epsilon|>\Delta} h_\epsilon \gamma_\epsilon.
\end{equation}
Here, $p_\epsilon$ and $h_\epsilon$ are the particle and hole components of the ABS wave function at the dot, respectively. The main technical trick of our calculation is to perform a rotation of the operators $\gamma_{E_\mathrm{A,0}}$ and $\gamma_{-E_\mathrm{A,0}}$ to a particle-hole basis such that
\begin{equation}
\label{eq:ph-operators-2}
d_\uparrow =\sqrt{\alpha}
\gamma_p+\sum_{|\epsilon|>\Delta}p_{\epsilon}\gamma_{\epsilon},\quad\quad\quad d_\downarrow^\dagger=\sqrt{\alpha}\gamma_h+\sum_{|\epsilon|>\Delta}h_{\epsilon}\gamma_{\epsilon},
\end{equation}
where fermionic operators $\gamma_{p/h}$ are defined as
\begin{equation}
\label{eq:rotation}
\gamma_p = \frac{1}{\sqrt{\alpha}}\left(p_{E_\mathrm{A,0}}\gamma_{E_\mathrm{A,0}}+p_{-E_\mathrm{A,0}}\gamma_{-E_\mathrm{A,0}}\right),\quad\quad\quad\gamma_h = \frac{1}{\sqrt{\alpha}}\left(h_{E_\mathrm{A,0}}\gamma_{E_\mathrm{A,0}}+h_{-E_\mathrm{A,0}}\gamma_{-E_\mathrm{A,0}}\right),
\end{equation}
and $\alpha$ is a normalization factor that ensures $\{\gamma_{p/h}, \gamma^\dagger_{p/h}\} = 1$ [this factor is similar for $\gamma_p$ and $\gamma_h$ due to particle-hole symmetry]. The particle and hole operators $\gamma_{p/h}$ are more convenient than operators $\gamma_{\pm E_\mathrm{A,0}}$ since the interaction Hamiltonian has a more concise form in terms of the former; this simplifies the calculation of the matrix elements of ${H}_\mathrm{int}$. 
We note that the normalization factor $\alpha$ can be expressed with the help of particle-hole symmetry as
\begin{equation}
    \alpha = |p_{E_\mathrm{A,0}}|^2 + |h_{E_\mathrm{A,0}}|^2 = |p_{-E_\mathrm{A,0}}|^2 + |h_{-E_\mathrm{A,0}}|^2.
\end{equation}
Thus, physically it describes the probability of finding a quasiparticle in the ABS at the dot rather than in the leads.

\subsubsection*{Corrections to energies of odd states}
We proceed by finding the correction to the energies of the odd states due to the presence of interaction. First, we note that the matrix element of $H_\mathrm{int}$ between different odd states vanishes since the Coulomb interaction conserves spin. The matrix elements between the odd and even states are zero as well, due to the conservation of fermion number parity. Therefore, to find the desired corrections to the odd states energies, it is enough to compute the expectation value of the interaction Hamiltonian in either of the odd states [the result is the same for the two odd states due to spin-rotation symmetry]. To evaluate this expectation value, it is convenient to express the odd states in terms of the particle and hole operators $\gamma_{p/h}$,
\begin{equation}
\label{eq:odd-p-h}
|1_{0,\uparrow}\rangle=\gamma_{p}^{\dagger}\gamma_{h}^{\dagger}\prod_{\epsilon<-\Delta} \gamma_\epsilon^\dagger  |\Omega\rangle,\quad\quad |1_{0,\downarrow}\rangle=\prod_{\epsilon<-\Delta} \gamma_\epsilon^\dagger  |\Omega\rangle.
\end{equation}
Substituting the decomposition of operators~\eqref{eq:ph-operators-2} into Eq.~\eqref{eq:Hint_app} and using Eq.~\eqref{eq:odd-p-h} we find for the projection of $H_\mathrm{int}$ on either of the odd states:
\begin{equation}
\label{eq:odd-state-correction}
\langle 1_{0,\sigma}|H_\mathrm{int}| 1_{0,\sigma} \rangle = - U\left( \sum_{\epsilon_{1},\epsilon_2<-\Delta}\left[|p_{\epsilon_{1}}|^2|h_{\epsilon_{2}}|^2
-h_{\epsilon_1}^{\star}p_{\epsilon_1}p_{\epsilon_2}^{\star}h_{\epsilon_2}\right]
-\frac{1}{2}\sum_{\epsilon<-\Delta}\left[|p_{\epsilon}|^2+|h_{\epsilon}|^2\right]
+\frac{1}{4}\right).
\end{equation}
Here, we used a set of relations that follow from particle-hole symmetry:
\begin{equation}
    \sum_{\epsilon>\Delta}p_{\epsilon}h_{\epsilon}^{\star} = -\sum_{\epsilon<-\Delta}p_{\epsilon}h_\epsilon^{\star},\quad\quad\quad \sum_{\epsilon>\Delta}|h_\epsilon|^2 = \sum_{\epsilon < -\Delta}|p_\epsilon|^2,\quad\quad\quad \sum_{\epsilon>\Delta}|p_\epsilon|^2 = \sum_{\epsilon < -\Delta} |h_\epsilon|^2,
\end{equation}
and the completeness relation for the single-particle wave-functions:
\begin{equation}\label{eq:compl_app}
\alpha+\sum_{\epsilon<-\Delta}|p_{\epsilon}|^2+\sum_{\epsilon<-\Delta}|h_{\epsilon}|^2=1.
\end{equation}
The sums over energies in Eq.~\eqref{eq:odd-state-correction} can be expressed in terms of the integrals of the Green's function of the dot. Direct comparison shows that
\begin{equation}
\sum_{\epsilon<-\Delta}
\begin{pmatrix}
\label{eq:pphh-vs-A}
p_\epsilon p_\epsilon^\star & p_\epsilon h_\epsilon^\star\\
p_\epsilon^\star h_\epsilon & h_\epsilon h_\epsilon^\star 
\end{pmatrix} = A,\quad \text{where}\quad A = \int_{-\infty}^{-\Delta} d\varepsilon \frac{i}{2\pi}\left[G^R_{dd}(\varepsilon)-G^A_{dd}(\varepsilon)\right].
\end{equation}
Using these relations in Eq.~\eqref{eq:odd-state-correction} we obtain Eq.~\eqref{eq:odd} of the main text.

Note that $\alpha = 1 - \mathrm{tr} A$ follows from Eqs.~\eqref{eq:compl_app} and \eqref{eq:pphh-vs-A}. The trace can be easily computed in terms of $E_\mathrm{A,0}$ by bending the integration contour in complex plane. This results in Eq.~\eqref{eq:alpha} of the main text.
 
\subsubsection*{Corrections to energies of even states}
The matrix elements of the interaction Hamiltonian in the even fermion parity sector are computed most easily in the basis of particle and hole states. These states are defined as
\begin{equation}
\label{eq:p-h-states}
|p\rangle=\gamma_{p}^{\dagger}|\mathcal{O}\rangle,\quad\quad\quad|h\rangle=\gamma_{h}^{\dagger}|\mathcal{O}\rangle
\end{equation}
[see Eq.~\eqref{eq:non-int-gs} for the definition of $|\mathcal{O}\rangle$]. The states $|p/h\rangle$ are directly related to the even states $|0_0 / 2_0\rangle$. Using Eq.~\eqref{eq:rotation} we obtain
\begin{equation}\label{eq:ph_app}
    |p\rangle = \frac{1}{\sqrt{\alpha}}\left(p_{-E_\mathrm{A,0}}^\star|0_0\rangle+p_{E_\mathrm{A,0}}^\star |2_0\rangle\right),\quad\quad\quad
    |h\rangle = \frac{1}{\sqrt{\alpha}}\left(h_{-E_\mathrm{A,0}}^\star|0_0\rangle + h_{E_\mathrm{A,0}}^\star|2_0\rangle\right). 
\end{equation}
The computation of the matrix elements of $H_{\rm int}$ between particle and hole states can be carried out similarly to how it was done for the odd states. This results in Eqs.~\eqref{eq:long-det} and \eqref{eq:even} of the main text.

\subsection{Low-energy expression for $E_\mathrm{A}$}
\label{app:int-low-en}
In this subsection, we use Eqs.~\eqref{eq:long-det} and \eqref{eq:even} to find the energy $E_\mathrm{A}$ in the limit $E_{\mathrm{A,0}}\ll \Delta$. This leads to Eqs.~\eqref{eq:EA} and \eqref{eq:renorm} of the main text. 
The limit $E_\mathrm{A,0}\ll \Delta$ is achieved in the weak coupling regime, $\Gamma,|\epsilon_g| \ll \Delta$ at any $\varphi$, and the strong coupling regime, $\Gamma\gtrsim\Delta$, provided that $|\Gamma_L - \Gamma_R|,|\epsilon_g|\ll \Delta + \Gamma$ and $|\varphi-\pi|\ll 1$.

According to Eq.~\eqref{eq:long-det}, $E_\mathrm{A}$ can be found as a solution of the following characteristic equation:
\begin{equation}
    \label{eq:long-long-det}
    \mathrm{det}\left[\varepsilon - 
    \frac{1}{1 + \frac{\Gamma}{\sqrt{\Delta^2 - E_\mathrm{A,0}^2}}}
    \begin{pmatrix}
    \epsilon_g & \frac{\Delta}{\sqrt{\Delta^2 - E_\mathrm{A,0}^2}}\sum_i\Gamma_i e^{i\varphi_i}\\
    \frac{\Delta}{\sqrt{\Delta^2 - E_\mathrm{A,0}^2}}\sum_i\Gamma_i e^{-i\varphi_i} & -\epsilon_g
    \end{pmatrix} - \alpha U
    \begin{pmatrix}
    \frac{A_{pp}-A_{hh}}{2} & A_{ph} \\
    A_{ph}^\star & -\frac{A_{pp}-A_{hh}}{2}
    \end{pmatrix}\right] = 0.
\end{equation}
Here, in the second term under the sign of $\det$,  $E_\mathrm{A,0}$ can be neglected in comparison to $\Delta$ in the square root factors. With a similar precision, $\alpha$ can be exchanged for its low-energy value, $\Delta/(\Delta+\Gamma)$. Finally, the elements of the matrix $A$ may be approximated by
\begin{equation}
\label{eq:f}
    \frac{A_{pp} - A_{hh}}{2} \approx \frac{\epsilon_g}{\Delta}f\left(\frac{\Gamma}{\Delta}\right),\quad f\left(x\right)  = - \frac{2x}{\pi}\int_{-\infty}^{-1}dz\frac{1}{\sqrt{z^{2}-1}}\frac{1}{z^{2}\left(1+\frac{x^{2}}{z^{2}-1}\right)^{2}} = -\frac{x}{\pi} \frac{2+x^2 - \frac{3 x \arccos(x)}{\sqrt{1-x^2}}}{(1-x^2)^2},
\end{equation}
\begin{equation}
\label{eq:g}
    A_{ph} \approx\frac{\Gamma_{R}e^{i\varphi_{R}}+\Gamma_{L}e^{i\varphi_{L}}}{\Delta}g\left(\frac{\Gamma}{\Delta}\right),\quad g(x) = \frac{1}{\pi}\int_{-\infty}^{-1} dz\frac{1}{\sqrt{z^2 - 1}}\frac{1-\frac{x^{2}}{z^{2}-1}}{z^{2}\left(1+\frac{x^{2}}{z^{2}-1}\right)^{2}} = \frac{1}{\pi}\frac{1+2x^2 - \frac{x (2 + x^2)\arccos(x)}{\sqrt{1-x^2}}}{(1-x^2)^2}.
\end{equation}
Substituting Eqs.~\eqref{eq:f} and \eqref{eq:g} with $\varphi_L = -\varphi_R = \varphi/2$ in Eq.~\eqref{eq:long-long-det} and solving the resulting simplified equation we obtain Eqs.~\eqref{eq:EA} and \eqref{eq:renorm} of the main text. Note that, at the first glance, functions $f(x)$ and $g(x)$ in Eqs.~\eqref{eq:f} and \eqref{eq:g} are divergent at $x = 1$ due to the vanishing of the denominators. This, however, is not the case because the numerators also vanish at $x = 1$ resulting in a smooth curve depicted in Fig.~\ref{fig:f-and-g}. In the weak coupling regime, $\Gamma,|\epsilon_g|\ll\Delta$, the results for $f$ and $g$, Eqs.~\eqref{eq:f} and \eqref{eq:g}, are reliable only to the leading order in $\Gamma / \Delta$. For $|\Gamma_L - \Gamma_R|,|\epsilon_g|\ll\Delta + \Gamma$ and $|\varphi-\pi|\ll 1$ the calculation of the matrix elements of $A$ is valid at arbitrary $\Gamma/\Delta$.

\subsection{Interaction corrections to energies $E[1]$ and $E_\mathrm{even}$ for weak coupling between the dot and the leads}
\label{app:int-weak-coup}
Here, we explicitly calculate energies $E[1]$ and $E_\mathrm{even}$ to the first order in $U / \Delta$ in the weak coupling limit, $\Gamma, |\epsilon_g| \ll \Delta$. The interaction correction to the energy of the odd states is given by
\begin{equation}
    \label{eq:odd-weak-coup}
    E[1] - E_0[1] \approx - U \det(A - 1/2)  \approx U\left[-\frac{1}{4}+\frac{\Gamma}{2\Delta} + \frac{3\Gamma\epsilon_g^2}{4\Delta^3}+\left(\frac{2}{\pi^2}-1\right)\frac{\Gamma_R\Gamma_L}{\Delta^2} \cos\varphi\right],
\end{equation}
where we retained only the leading terms that determine the $\epsilon_g$- and $\varphi$- dependence of $E[1] - E_0[1]$. In particular, we suppressed small terms $\sim U\epsilon_g^2 \Gamma_R\Gamma_L\cos\varphi/\Delta^4$ that depend both on $\epsilon_g$ and on $\varphi$. In principle, such terms are important for the careful calculation of the adiabatic contribution to $\chi_{QI}$ and $\chi_{IQ}$. However, capturing them analytically is beyond the scope of the manuscript.

With a similar precision we find
\begin{equation}
    \label{eq:even-weak-coup}
    E_\mathrm{even} - E_0[1] \approx - U \det(A - 1/2) + U\frac{\alpha^2}{2} \approx U\left[\frac{1}{4}-\frac{\Gamma}{2\Delta} - \frac{3\Gamma\epsilon_g^2}{4\Delta^3}+\left(\frac{2}{\pi^2}+1\right)\frac{\Gamma_R\Gamma_L}{\Delta^2} \cos\varphi\right].
\end{equation}

\section{Low-energy Hamiltonian}
\label{sec:app-low-en-ham}
In this Appendix, we derive the low-energy Hamiltonian governing the dynamics of the ABS in the even fermion parity sector [Eq.~\eqref{eq:low-en} of the main text]. To do that, we project the full Hamiltonian of the system onto the low-energy subspace, and then employ the adiabatic approximation. The low-energy subspace is formed by the discrete states $|0\rangle$ and $|2\rangle$, whose energies are denoted by $E[0/2] = E_{\rm even} \mp E_{\rm A}$. The low-energy regime of $E_{\rm A} \ll \Delta$ is reached in two cases: (i) in the weak coupling limit, $|\epsilon_g|, \Gamma \ll \Delta$ at arbitrary phase bias $\varphi$, and (ii) in the strong coupling limit, $\Gamma \gtrsim \Delta$, provided $|\varphi - \pi| \ll 1$ and $|\Gamma_L - \Gamma_R|, |\epsilon_g| \ll \Delta + \Gamma$. We assume below that either of the two conditions is fulfilled. 

We first focus on the case in which the Coulomb interaction is absent, $U = 0$; we discuss the modifications arising due to $U \neq 0$ at the end of the section.
Let us consider the many-body Hamiltonian $H[\varphi_i, \epsilon_g]$ [Eq.~\eqref{eq:model} with $U = 0$], in which parameters $\varphi_i \equiv \varphi_i(t)$ and $\epsilon_g \equiv \epsilon_g(t)$ depend on time.
We assume that the dynamics of $\varphi_i(t)$ and $\epsilon_g(t)$ is sufficiently slow --- \textit{i.e.}, the associated frequency scale $\hbar\omega \ll \Delta$. At the same time, we allow $\hbar\omega$ to be comparable to $E_{\rm A} \ll \Delta$, which makes it important to account for possible transitions between the states of the low-energy subspace.

The wave-function solving the time-dependent Schr\"odinger equation can be approximated by
\begin{equation}\label{eq:approx_wf_app}
    |\psi(t)\rangle \approx c_p(t)|p(t)\rangle + c_h(t)|h(t)\rangle.
\end{equation}
Here the particle and hole states $|p/h\rangle$ were introduced in Eq.~\eqref{eq:ph_app}; they depend on time parametrically due to the time-dependence of $\varphi_i(t)$  and $\epsilon_g(t)$. This parametric dependence is in fact weak, which makes the particle-hole basis convenient for the derivation. Finding an effective Hamiltonian that would describe the evolution of amplitudes $c_p(t)$ and $c_h(t)$ is the main goal of this Appendix. The approximation \eqref{eq:approx_wf_app} is in the spirit of a usual adiabatic approximation extended to a two-level system.

Substituting the decomposition \eqref{eq:approx_wf_app} into the time-dependent Schr\"odinger equation, we obtain an equation for the evolution of $C(t) = (c_p(t),\,c_h(t))^T$:
\begin{equation}
    i\hbar\partial_t C(t) = H_\mathrm{even}^{\mathrm{(le)}}\,C(t).
\end{equation}
The $2\times 2$ matrix $H_\mathrm{even}^{\mathrm{(le)}}$ plays the role of an effective Hamiltonian; it is given by
\begin{equation}\label{eq:le-ham_gen_app}
    H_\mathrm{even}^{\mathrm{(le)}} = E_{\rm even} + {\cal H} + \Theta.
\end{equation}
Here
\begin{equation}
    {\cal H}_{\mu\nu} = \langle \mu(t) | H[\varphi_i(t), \epsilon_g(t)] |\nu(t) \rangle - E_{\rm even}\delta_{\mu\nu}, \quad \Theta_{\mu\nu} = -i\hbar \langle \mu(t) | {\partial_t}| \nu(t)\rangle,\quad \mu,\nu \in \{ p, h \},
\end{equation}
and $E_{\rm even}$ is a $c$-number term which --- in the absence of Coulomb interaction --- is related to the continuum energy, $E_{\rm even} = E_{\rm cont}(\epsilon_g(t),\varphi(t))$ [see Eq.~\eqref{eq:cont}].
The matrix ${\cal H}$ can be found straightforwardly using the definition \eqref{eq:ph_app} of particle and hole states. We obtain
\begin{equation}\label{eq:H_app}
    {\cal H} = \frac{1}{1 + \frac{\Gamma}{\sqrt{\Delta^2 - E_\mathrm{A,0}^2(t)}}}
    \begin{pmatrix}
    \epsilon_g(t) & \frac{\Delta \sum_i\Gamma_i e^{i\varphi_i(t)}}{\sqrt{\Delta^2 - E_\mathrm{A,0}^2(t)}} \\
    \frac{\Delta \sum_i\Gamma_i e^{-i\varphi_i(t)}}{\sqrt{\Delta^2 - E_\mathrm{A,0}^2(t)}}  & -\epsilon_g(t)
    \end{pmatrix} \approx \frac{\Delta}{\Delta + \Gamma}
    \begin{pmatrix}
    \epsilon_g(t) & \sum_i\Gamma_i e^{i\varphi_i(t)}\\
    \sum_i\Gamma_i e^{-i\varphi_i (t)} & -\epsilon_g(t)
    \end{pmatrix},
\end{equation}
where in the latter equality we neglected $E_{\rm A,0} \ll \Delta$.

Next, $\Theta$ in Eq.~\eqref{eq:le-ham_gen_app} is the matrix of Berry connection. It stems from the parametric dependence of $|p(t)\rangle$ and $|h(t)\rangle$ on time. To find $\Theta$, it is convenient to use the many-body representation for the states: $|p/h\rangle = \gamma^\dagger_{p/h}|{\cal O}\rangle$. This representation allows us to rewrite $\Theta$ as
\begin{equation}\label{eq:theta_app}
    \Theta_{\mu\nu} = -i\hbar\langle {\cal O} | \gamma_{\mu} \dot{\gamma}^\dagger_{\nu} | {\cal O}\rangle -i\hbar\delta_{\mu\nu} \langle {\cal O} | \dot{\cal O}\rangle.
\end{equation}
Here the second term is $\propto \delta_{\mu\nu}$ and thus does not influence the dynamics of the system [\textit{e.g.}, it has no effect on the response functions]; we omit it in what follows.   To find the first term, $-i\hbar\langle {\cal O} | \gamma_{\mu} \dot{\gamma}^\dagger_{\nu} | {\cal O}\rangle$, it is convenient to expand $\gamma_{p}$ and $\gamma_h$ into the electron field operators. The expansion reads:
\begin{equation}\label{eq:decomp_app}
        \begin{pmatrix}
        \gamma_p\\\gamma_h
        \end{pmatrix}
        =
        \sqrt{\alpha}
        \begin{pmatrix}
        d_{\uparrow}\\
        d_{\downarrow}^\dagger
        \end{pmatrix}
        + \sum_{i, k} \Pi^\dagger_{i,k}
        \begin{pmatrix}
        \psi_{\uparrow, i,k}\\
        \psi^\dagger_{\downarrow, i,-k}
        \end{pmatrix},
\end{equation}
where $d_\sigma$ is the annihilation operator for an electron at the dot with spin $\sigma =\,\uparrow$ or $\downarrow$, and $\psi_{\sigma, i,k}$ is the annihilation operator for an electron in the lead $i\in \{L, R\}$ with momentum $k$ and spin projection $\sigma$. Parameter $\alpha$ is defined in Eqs.~\eqref{eq:alpha_int}, \eqref{eq:alpha}. Finally, $\Pi_{i,k}$ is a $2 \times 2$ matrix that is defined in terms of the ABS wave-functions as
\begin{equation}
    \Pi_{i,k} = \frac{1}{\sqrt{\alpha}} \left(\Phi_{E_{\rm A, 0},i,k},\, \Phi_{-E_{\rm A, 0}, i, k}\right) 
    \begin{pmatrix}
        p_{E_{\rm A, 0}} & p_{-E_{\rm A, 0}}\\
        h_{E_{\rm A, 0}} & h_{-E_{\rm A, 0}}
    \end{pmatrix}.
\end{equation}
Here, $p_{\pm E_{\rm A, 0}}, h_{\pm E_{\rm A, 0}}$ are the components of the wave-function at the dot [see Eq.~\eqref{eq:ph-operators-1}], and $\Phi_{\pm E_{\rm A, 0},i,k}$ are the $2 \times 1$ spinors describing the components of the ABS wave-functions in the lead $i$. Using the decomposition \eqref{eq:decomp_app} together with the normalization condition $\alpha + \sum_{i,k} \Pi_{i,k}^\dagger \Pi_{i,k} = 1$ [where $\alpha$ and $1$ are proportional to the $2 \times 2$ identity matrices] in Eq.~\eqref{eq:theta_app} we may represent the Berry connection $\Theta$ as
\begin{equation}\label{eq:theta_app2}
    \Theta = -\frac{i\hbar}{2}\sum_{i,k}\left[\Pi_{i,k}^{\dagger}\dot{\Pi}_{i,k}-\dot{\Pi}_{i,k}^{\dagger}\Pi_{i,k} \right].
\end{equation}
The matrix $\Pi_{i,k}$ can be found using the Schr\"odinger equation for the ABS wave-functions. We obtain
\begin{equation}\label{eq:pi_app}
    \Pi_{i, k} = \frac{1}{\sqrt{V}}\frac{\sqrt{\alpha}\, T_i}{\Delta^{2}-E_{\rm A,0}^{2}+\xi_{k}^{2}}\left[\xi_{k}\tau_{z}-\Delta e^{i\varphi_{i}\tau_{z}/2}\tau_{x}e^{-i\varphi_{i}\tau_{z}/2}+{\cal H}\right],
\end{equation}
where $\tau_{x,y,z}$ are Pauli matrices in the Nambu space, $T_i = t_{i}\tau_{z}e^{-i\varphi_{i}\tau_{z}/2}$, and $V$ is the volume of the lead. The third term in brackets is $\sim E_{\rm A,0}$ and may be neglected in comparison with the first two terms $\sim \Delta$ at low energies $E_{\rm A,0} \ll \Delta$~\footnote{It can be shown by a straightforward-yet-tedious calculation that retaining ${\cal H}$ in Eq.~\eqref{eq:pi_app} would only produce corrections to the Berry connection $\Theta$ whose influence on the dynamics of the ABS (\textit{e.g.}, on the transition matrix elements) is suppressed by a small parameter $E_{\rm A,0}^2 / \Delta^2 \ll 1$.}. Then, we find
\begin{equation}
        \Pi_{i, k} \approx \frac{1}{\sqrt{V}} \sqrt{\frac{\Delta}{\Delta + \Gamma}}\frac{T_i}{\Delta^{2}+\xi_{k}^{2}}\left[\xi_{k}\tau_{z}-\Delta e^{i\varphi_{i}\tau_{z}/2}\tau_{x}e^{-i\varphi_{i}\tau_{z}/2}\right],
\end{equation}
where we also disregarded $E_{\rm A,0}$ is the denominator and used $\alpha \approx \Delta / (\Delta + \Gamma)$. Substituting this expression in Eq.~\eqref{eq:theta_app2} we obtain
\begin{equation}\label{eq:connection_app}
    \Theta \approx -\tau_z \, \frac{\Gamma_R \hbar\dot{\varphi}_R + \Gamma_L \hbar\dot{\varphi}_L}{2(\Delta + \Gamma)}.
\end{equation}

Combining Eqs.~\eqref{eq:le-ham_gen_app}, \eqref{eq:H_app}, and \eqref{eq:connection_app}, we arrive to a final expression for the low-energy Hamiltonian in the absence of Coulomb interaction:
\begin{equation}\label{eq:low-en-ham_final_app}
    H_\mathrm{even}^{\mathrm{(le)}} = E_{\rm even} + \frac{\Delta}{\Delta + \Gamma}
    \begin{pmatrix}
    \epsilon_g(t) & \sum_i\Gamma_i e^{i\varphi_i (t)}\\
    \sum_i\Gamma_i e^{-i\varphi_i(t)} & -\epsilon_g(t)
    \end{pmatrix} 
    - \frac{\Gamma_R \hbar\dot{\varphi}_R(t) + \Gamma_L \hbar\dot{\varphi}_L(t)}{2(\Delta + \Gamma)}
    \begin{pmatrix}
    1 & 0 \\
    0 & -1
    \end{pmatrix}.
\end{equation}
The Hamiltonian $H_\mathrm{even}^{\mathrm{(le)}}$ is evidently consistent with the gauge-invariance. The gauge-invariance means that the physics should not be affected by a common shift $V_{\rm sh}(t)$ of all electric potentials,
\begin{equation}\label{eq:gauge1_app}
    \epsilon_g(t) \rightarrow \epsilon_g(t) - e V_{\rm sh}(t),\quad\quad \hbar\dot{\varphi}_{R/L}(t) \rightarrow \hbar\dot{\varphi}_{R/L}(t) + 2e V_{\rm sh}(t). 
\end{equation}
Indeed, as can be easily checked, such a common shift may be compensated by a unitary transformation~\footnote{The gauge-invariance of $E_{\rm even}$ is ensured by a contribution to the Berry connection $-i\hbar \langle {\cal O}|\dot{{\cal O}}\rangle$, which we omitted in Eq.~\eqref{eq:low-en-ham_final_app}.}
\begin{equation}\label{eq:gauge2_app}
    H_\mathrm{even}^{\mathrm{(le)}} \rightarrow {\cal U}H_\mathrm{even}^{\mathrm{(le)}} {\cal U}^\dagger - i\hbar\,{\cal U} \dot{\cal{U}}^\dagger, \quad \text{where} \quad {\cal U} = \exp \Bigl[-i\tau_z \frac{1}{\hbar}\int^t e V_{\rm sh}(t^\prime) dt^\prime \Bigr].
\end{equation}

The next step is to account for a weak Coulomb interaction, $U \ll \Delta + \Gamma$. The interaction leads to the renormalization of the parameters of the low-energy Hamiltonian $H_\mathrm{even}^{\mathrm{(le)}}$. The renormalizations may be accounted for by combining the perturbative approach of Appendix~\ref{app:int} with the requirement of the gauge-invariance, as detailed below.

To start with, it is again convenient to represent the low-energy Hamiltonian as $H_\mathrm{even}^{\mathrm{(le)}} = E_{\rm even} + {\cal H} + \Theta$, where ${\cal H}_{\mu\nu} = \langle \mu| H |\nu\rangle - E_{\rm even}$ and $\Theta_{\mu\nu} = -i\hbar \langle \mu | \partial_t | \nu \rangle$. To find $E_{\rm even}$ and ${\cal H}$, we project the full many-body Hamiltonian $H$ (including the Coulomb interaction part) onto particle and hole states which depend on time parametrically. The projection is carried out similarly to how it was done in Appendix~\ref{app:int}. We obtain $E_{\rm even}$ given by Eq.~\eqref{eq:even} and ${\cal H}$ given by
\begin{equation}\label{eq:low-en-ham_int_app}
    {\cal H} \approx \frac{\Delta}{\Delta + \Gamma}
    \begin{pmatrix}
    \epsilon_g(t) \left[1 + \frac{U}{\Delta} f(\frac{\Gamma}{\Delta})\right] & \sum_i \Gamma_i e^{i\varphi_i(t)} \left[1 + \frac{U}{\Delta} g(\frac{\Gamma}{\Delta})\right]\\
    \sum_i \Gamma_i e^{-i\varphi_i(t)} \left[1 + \frac{U}{\Delta} g(\frac{\Gamma}{\Delta})\right] & - \epsilon_g(t) \left[1 + \frac{U}{\Delta} f(\frac{\Gamma}{\Delta})\right]
    \end{pmatrix}.
\end{equation}
Here dimensionless functions $f$ and $g$ are defined in Eqs.~\eqref{eq:f} and \eqref{eq:g}, respectively.

To understand how expression \eqref{eq:theta_app} for the Berry connection $\Theta$ gets renormalized by $U \neq 0$, we require the low-energy Hamiltonian $H^{\rm (le)}_{\rm even} = E_{\rm even} + {\cal H} + \Theta$ (with ${\cal H}$ given by Eq.~\eqref{eq:low-en-ham_int_app}) to be consistent with the gauge-invariance [see Eqs.~\eqref{eq:gauge1_app} and \eqref{eq:gauge2_app}]. This leads~to 
\begin{equation}\label{eq:theta_int_app}
    \Theta \approx -\tau_z \, \frac{\Gamma_R \hbar\dot{\varphi}_R + \Gamma_L \hbar\dot{\varphi}_L}{2(\Delta + \Gamma)} \left[1 - \frac{U}{\Gamma} f\left(\frac{\Gamma}{\Delta}\right)\right]
\end{equation}
In the main text, we focus on a particular gauge in which $\varphi_L(t) = -\varphi_R(t) = \varphi(t) / 2$. In this gauge, we find by combining Eqs.~\eqref{eq:low-en-ham_int_app} and \eqref{eq:theta_int_app}:
\begin{equation}\label{eq:low-en-ham_int_app2}
    H_\mathrm{even}^{\mathrm{(le)}} \approx E_{\rm even} + \frac{\Delta}{\Delta + \Gamma}
    \begin{pmatrix}
    \tilde{\epsilon}_g(t) & \tilde{\gamma}[\varphi(t)] \\
    \tilde{\gamma}^\star[\varphi(t)]  & - \tilde{\epsilon}_g(t)
    \end{pmatrix}
    - \frac{\delta \Gamma\,\hbar\dot{\varphi}(t)}{4(\Delta + \Gamma)}\left[1 - \frac{U}{\Gamma} f\left(\frac{\Gamma}{\Delta}\right)\right] \begin{pmatrix}
        1 & 0 \\
        0 & -1
    \end{pmatrix},
\end{equation}
where $\delta \Gamma = \Gamma_L - \Gamma_R$ and
\begin{equation}
    \tilde{\epsilon}_g(t) = \epsilon_g(t) \left[1 + \frac{U}{\Delta} f\left(\frac{\Gamma}{\Delta}\right)\right],\quad\quad \tilde{\gamma}[\varphi(t)] =  \left(\Gamma \cos \frac{\varphi(t)}{2} + i\delta\Gamma \sin \frac{\varphi(t)}{2}\right) \left[1 + \frac{U}{\Delta} g\left(\frac{\Gamma}{\Delta}\right)\right].
\end{equation}
Finally, we note that in the considered gauge the Berry connection term in Eq.~\eqref{eq:low-en-ham_int_app2} is small: at most, it produces relative corrections $\sim E_{\rm A} / \Delta$ to the transition matrix elements. The renormalization of the Berry connection due to $U \ll \Delta + \Gamma$ thus has a very weak influence on the dynamics of the ABS (which is controlled by a small parameter $\sim E_{\rm A} U / \Delta^2 \ll 1$). Neglecting the renormalization, we arrive to Eq.~\eqref{eq:low-en} of the main text.

\section{Linear response functions}
\label{sec:app-resp-fun}
\subsection{General expression for the linear response functions}
\label{app:general_lin_resp}
In this section, we derive Eq.~\eqref{eq:response} of the main text. We start with a general linear response relation,
\begin{equation}
    \chi_{AB}[\omega, n] = \langle \partial_b \hat{A}\rangle + \chi^\mathrm{K}_{AB}[\omega, n],\quad \chi^\mathrm{K}_{AB}[\omega,n] = -i\int_{0}^{\infty}dt e^{i\omega t}\langle[\hat{A}(t),\hat{B}(0)]\rangle,
\end{equation}
that includes the diamagnetic term $\langle \partial_b \hat{A}\rangle$ and the Kubo term $\chi^\mathrm{K}_{AB}[\omega,n]$. This relation can be expressed identically as
\begin{equation}
    \chi_{AB}[\omega, n] = \langle\partial_b \hat{A} \rangle + \chi_{AB}^\mathrm{K}[0, n] + \delta\chi_{AB}[\omega, n],\quad \text{where}\quad \delta\chi_{AB}[\omega, n] = \chi_{AB}^\mathrm{K}[\omega, n] - \chi_{AB}^\mathrm{K}[0, n].  
\end{equation}
By definition, the dynamic part of the response function $\delta\chi^\mathrm{K}_{AB}[\omega,n]$ vanishes at zero frequency. The remaining part, $\langle\partial_b \hat{A} \rangle + \chi_{AB}^\mathrm{K}[0, n]$, in turn describes the zero-frequency response. We may simplify the latter part using the fact that at $\omega = 0$ the system follows the applied drives adiabatically. The adiabaticity implies
\begin{equation}
    \chi_{AB}[0, n] = \partial_{b}\langle \hat{A} \rangle = \partial_{a} \partial_b E[n],
\end{equation}
where we used $\langle \hat{A} \rangle = \langle \partial_a H\rangle = \partial_a E[n]$.  Thus we identify
\begin{equation}
     \langle\partial_b \hat{A} \rangle + \chi_{AB}^\mathrm{K}[0, n] = \chi_{AB}[0, n] \equiv\partial_{a}\partial_{b} E[n],
\end{equation}
which leads to Eq.~\eqref{eq:response} of the main text.
\begin{figure*}[t]
  \begin{center}
    \includegraphics[scale=1]{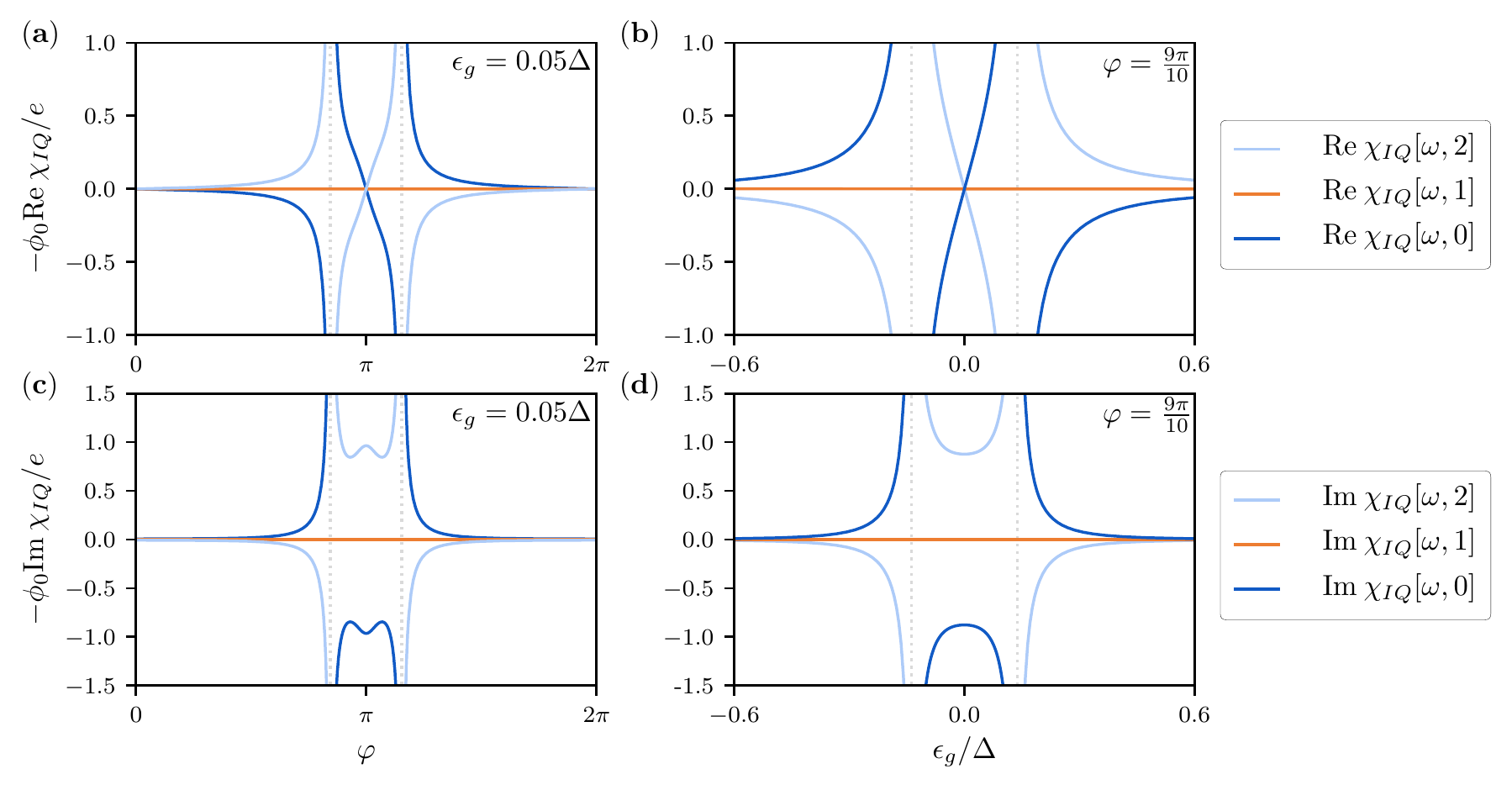}
    \caption{The response function $\chi_{IQ}$ in states $|0\rangle$, $|1_\sigma\rangle$, $|2\rangle$ with different number of quasiparticles at the ABS. $\mathrm{Re}\,\chi_{IQ}$ is plotted as a function of $\varphi$ in panel (a) [for $\epsilon_g = 0.05\Delta$] and as a function of $\epsilon_g$ in panel (b) [for $\varphi = 9\pi/10$]; $\mathrm{Im}\,\chi_{IQ}$ is plotted as a function of $\varphi$ in panel (c) [for $\epsilon_g = 0.05\Delta$] and as a function of $\epsilon_g$ in panel (d) [for $\varphi = 9\pi/10$]. The plots are produced using Eqs.~\eqref{eq:response}, \eqref{eq:0=-2}, \eqref{eq:chi-IQ}, for parameters $\Gamma_L = 0.3 \Delta$, $\Gamma_R = 0.35 \Delta$, $U = 0.35 \Delta$, and $\hbar\omega = 0.21 \Delta$ [these parameters are similar to that in Fig.~\ref{fig:response}]. The response functions in states $|0\rangle$ and $|2\rangle$ diverge when the frequency is in resonance with the transition between $|0\rangle$ and $|2\rangle$, \textit{i.e.}, when $\hbar\omega = 2E_A$ (vertical dashed lines in the plots). Generally, the response function $\chi_{IQ}$ in the odd states is small compared to that in the even states. We note that, away from resonances, both ${\rm Re}\,\chi_{IQ}$ and ${\rm Im}\,\chi_{IQ}$ describe the non-dissipative response. The dissipative part of the response functions --- which is present at resonances only --- is not shown in the plot.}
    \label{fig:response-app}
  \end{center}
\end{figure*}

\subsection{Off-diagonal components of the response function\label{app:iq-and-qi}}
In this Appendix, we present the plots of phase- and gate voltage- dependence of $\chi_{IQ}[n, \omega]$ for the parameters $\Gamma_L, \Gamma_R, U$ and $\omega$ similar to those used in Fig.~\ref{fig:response}. Non-dissipative parts of  $\chi_{IQ}$ and $\chi_{QI}$ are related by the complex conjugation and thus we do not consider the latter separately. In contrast to $\chi_{QQ}$ and $\chi_{II}$, non-dissipative part of  $\chi_{IQ}$ has both real and imaginary components. We plot these components separately in Fig.~\ref{fig:response-app}. Note that $\chi_{IQ}$ vanishes if $\epsilon_g = 0$ or $\varphi = \pi$ (as a consequence of particle-hole and time-reversal symmetries, respectively [see discussion after Eq.~\eqref{eq:inv-symm}]). Thus, we plot the phase dependence for $\epsilon_g = 0.05\Delta$ and the gate voltage dependence for $\varphi = 9 \pi / 10$.

\subsection{Full expressions for dynamic parts of the response functions at small frequencies\label{app:full-expr}}
Here, we present expressions for the dynamic parts of the response functions without neglecting the small phase factor in Eq.~\eqref{eq:z} (as was done in the main text). We only present expressions for $\delta\chi_{AB}$ in the state $|0\rangle$. In the state $|2\rangle$ the dynamic part of the response function can be approximately recovered as $\delta\chi_{IQ}[\omega, 2]= -\delta\chi_{IQ}[\omega,0]$. The dynamic part of the response function in the odd states is small at $\hbar\omega\ll\Delta$.

Equation \eqref{eq:chi-QQ} for $\delta\chi_{QQ}$ remains unaltered when the phase factor is taken into the account. For $\delta\chi_{II}$ at $\hbar\omega\ll\Delta$ we get
\begin{equation}
    \delta\chi_{II}[\omega, 0] = -\frac{\phi_0^{-2}}{E_\mathrm{A}} \frac{\hbar^2\omega^2}{4E_\mathrm{A}^2-(\hbar\omega+i0)^2}\frac{1}{|\tilde{\gamma}|^2}\Bigg[\tilde{\epsilon}_g^2 (\partial_\varphi E_\mathrm{A})^2   +\frac{1}{4}\left(\frac{\Delta}{\Delta+\Gamma}\right)^2\left(1+g \frac{U}{\Delta}\right)^4\delta\Gamma^2\left(\Gamma - \frac{|\gamma|^2}{\Delta+\Gamma}\right)^2\Bigg],
\end{equation}
where $|\gamma|^2 = \Gamma^2 - 4\Gamma_R\Gamma_L \sin^2(\varphi/2)$. For $\delta\chi_{IQ}$ we obtain at $\hbar\omega \ll \Delta$
\begin{equation}
\label{eq:chi-iq-app}
    \delta\chi_{IQ}[\omega, 0] = \frac{e\phi_0^{-1}}{4 E_\mathrm{A}^2 - (\hbar\omega + i0)^2}\Bigg[\hbar^2\omega^2 \partial_{\epsilon_g}\partial_{\varphi} E_{\mathrm{A}}+ \left(\frac{\Delta}{\Delta + \Gamma}\right)^3\frac{i\hbar\omega}{E_{\mathrm{A}}}\left(\Gamma - \frac{|\gamma|^2}{\Delta+\Gamma}\right)\delta\Gamma\left(1+f \frac{U}{\Delta}\right)\left(1+g \frac{U}{\Delta}\right)^2\Bigg].
\end{equation}
Response function $\delta\chi_{QI}$ can be obtained from Eq.~\eqref{eq:chi-iq-app} by conjugating the expression in the square brackets.
\subsection{Exact evaluation of the linear-response functions in the absence of interaction}
\label{sec:app-exact}
In this Appendix, we provide the exact expressions for linear response functions in the discrete states $|0_0\rangle$, $|1_{0,\sigma}\rangle$, and $|2_0\rangle$ in the absence of Coulomb interaction. We also clarify why in the non-interacting  case the linear response functions satisfy the occupation rule ($\chi[\omega, 0]+\chi[\omega, 2])/2 = \chi[\omega, 1]$. Again, we use units with $\hbar = 1$.

We start with a general expression for the linear response functions. As was shown in Appendix~\ref{app:general_lin_resp}, 
\begin{gather}
    \chi_{AB}[\omega, n] = \partial_a \partial_b E_0[n] + \delta\chi_{AB}[\omega,n],\\
   \delta\chi_{AB}[\omega,n] = \chi^\mathrm{K}_{AB}[\omega,n] - \chi^\mathrm{K}_{AB}[0,n],\quad \chi^\mathrm{K}_{AB}[\omega,n] = -i\int_{0}^{\infty}dt e^{i\omega t}\langle[\hat{A}(t),\hat{B}(0)]\rangle.
\end{gather}
Here, $A$ and $B$ stand for either current or charge, $a$ and $b$ are the corresponding drive variables ($V_g$ corresponds to $Q$ and $\phi$ corresponds to $I$), and the average in $\chi_{AB}^K$ is taken over the discrete state $|n_0\rangle$ (with $n=0,1$ or $2$).
The adiabatic part of the response function, $\partial_a\partial_b E_0[n]$, can be calculated using the exact expression for the energies of the discrete states, cf.~Eq.~\eqref{eq:unpert}. To calculate the dynamic part, $\delta\chi_{AB}$, it is convenient to introduce the single-particle representations of the current and charge operators $\mathcal{I}$ and $\mathcal{Q}$. These objects are related to the corresponding many-body operators $\hat{I}$ and $\hat{Q}$ through
\begin{equation}
\hat{I} =\eta^{\dagger} \mathcal{I} \eta,\quad \hat{Q} =\eta^{\dagger} \mathcal{Q} \eta,\quad\text{where}\quad\eta = \begin{pmatrix}
D\\
\Psi_L(r=0)\\
\Psi_R(r=0)
\end{pmatrix},
\end{equation}
and are given explicitly by
\begin{equation}
\label{eq:single-particle-ops}
\mathcal{I} = -\frac{e}{2}\tau_{z}\begin{pmatrix}
0 & -it_{L}e^{i\varphi_{L}\tau_{z}/2} & it_{R}e^{i\varphi_{R}\tau_{z}/2}\\
it_{L}e^{-i\varphi_{L}\tau_{z}/2} & 0 & 0\\
-it_{R}e^{-i\varphi_{R}\tau_{z}/2} & 0 & 0
\end{pmatrix}, \quad\quad
\mathcal{Q} = - e \tau_z
\begin{pmatrix}
1 & 0 & 0\\
0 & 0 & 0\\
0 & 0 & 0
\end{pmatrix}.
\end{equation}
We will now use the Wick's theorem to express the average $\langle [\hat{A}(t), \hat{B}(0)] \rangle$ in terms of the single-particle Green's functions and matrix elements of the operators ${\cal I}$ and ${\cal Q}$ [see Eq.~\eqref{eq:single-particle-ops}]. To do that, we first introduce the relevant Green's functions:
\begin{equation}
\label{eq:pmmp}
\mathcal{G}_{\mu\nu}^{+-}(t)=-i\langle\eta_{\mu}(t)\eta_{\nu}^{\dagger}(0)\rangle,\quad\quad\quad \mathcal{G}_{\mu\nu}^{-+}(t)=i\langle\eta_{\nu}^{\dagger}(0)\eta_{\mu}(t)\rangle,
\end{equation}
where $\mu, \nu$ are the indexes in the Nambu space. In terms of these Green's functions the response functions can be expressed as
\begin{equation}
    \label{eq:kubo-vs-gf}
    \chi^\mathrm{K}_{AB}[\omega, n] = -i\int_{0}^{\infty}dte^{i\omega t}\mathrm{Tr}\left[\mathcal{A}\mathcal{G}^{+-}(t)\mathcal{B}\mathcal{G}^{-+}(-t)-\mathcal{A}\mathcal{G}^{-+}(t)\mathcal{B}\mathcal{G}^{+-}(-t)\right],
\end{equation}
where $\mathcal{A}$ and $\mathcal{B}$ are the single-particle versions of the operators $\hat{A}$ and $\hat{B}$, respectively, and the trace is taken over the matrix indices both in the dot/lead subspace and in the Nambu subspace. Note that the information about the state of the system --- $|0_0\rangle$, $|1_{0,\sigma}\rangle$, or $|2_0\rangle$ --- is encoded in ${\cal G}^{+-}$ and ${\cal G}^{-+}$. Next, it is convenient to relate ${\cal G}^{+-}$ and ${\cal G}^{-+}$ to the retarded, advanced, and Keldysh Green's functions,
\begin{equation}
\label{eq:RAK}
\mathcal{G}_{\mu\nu}^{R/A}(t)=\mp i\theta(\pm t)\langle\{\eta_{\mu}(t),\eta_{\nu}^{\dagger}(0)\}\rangle,\quad \mathcal{G}_{\mu\nu}^{K}(t)=\mathcal{G}_{\mu\nu}^{+-}(t)+\mathcal{G}_{\mu\nu}^{-+}(t).
\end{equation}
The relations may be summarized as
\begin{equation}
\label{eq:ULRAK}
\begin{gathered}
\mathcal{G}^{+-}(t>0)=\frac{\mathcal{G}^{K}(t)+\mathcal{G}^{R}(t)}{2},\quad \mathcal{G}^{-+}(t>0)=\frac{\mathcal{G}^{K}(t)-\mathcal{G}^{R}(t)}{2},\\
\mathcal{G}^{+-}(t<0)=\frac{\mathcal{G}^{K}(t)-\mathcal{G}^{A}(t)}{2},\quad \mathcal{G}^{-+}(t<0)=\frac{\mathcal{G}^{K}(t)+\mathcal{G}^{A}(t)}{2}.
\end{gathered}
\end{equation}
Representation of Eq.~\eqref{eq:ULRAK} is convenient, because the retarded and advanced Green's functions are agnostic to the state of the system. The information about the latter is solely contained in the Keldysh Green's function.  Substituting Eq.~\eqref{eq:ULRAK} into Eq.~\eqref{eq:kubo-vs-gf} we obtain
\begin{equation}
\label{eq:kubo-vs-keldysh}
\chi_{AB}^\mathrm{K}[\omega, n] = -\frac{i}{2}\int_{0}^{\infty}dt e^{i\omega t}\mathrm{Tr}\left[\mathcal{A}\mathcal{G}^{R}(t)\mathcal{B}\mathcal{G}^{K}(-t)+\mathcal{A}\mathcal{G}^{K}(t)\mathcal{B}\mathcal{G}^{A}(-t)\right]
\end{equation}
As the next step, we transfer the latter equation to the energy domain and use Kramers-Kronig relations
\begin{equation}
\label{eq:GRA}
\mathcal{G}^{R}(\hbar\omega+\varepsilon_{1})=-\int_{-\infty}^{\infty}\frac{d\varepsilon_{2}}{2\pi i}\frac{\mathcal{G}^{R}(\varepsilon_{2})-\mathcal{G}^{A}(\varepsilon_{2})}{\hbar\omega+\varepsilon_{1}-\varepsilon_{2}+i0},\quad \mathcal{G}^{A}(\varepsilon_{1}-\hbar\omega)=-\int_{-\infty}^{\infty}\frac{d\varepsilon_{2}}{2\pi i}\frac{\mathcal{G}^{R}(\varepsilon_{2})-\mathcal{G}^{A}(\varepsilon_{2})}{\varepsilon_{1}-\hbar\omega-\varepsilon_{2}-i0},
\end{equation}
as well as
\begin{equation}
    \label{eq:keldysh-vs-n}
    \mathcal{G}^K(\varepsilon) = (1-2n(\varepsilon))\left[\mathcal{G}^{R}(\varepsilon)-\mathcal{G}^{A}(\varepsilon)\right].
\end{equation}
In the Eq.~\eqref{eq:keldysh-vs-n} $n(\varepsilon)$ is the distribution function. For the considered discrete states $n(\varepsilon>\Delta) = 0$ and $n(\varepsilon<-\Delta) = 1$. State $|0_0\rangle$ is determined by $n(-E_\mathrm{A,0}) = 1$ and $n(E_\mathrm{A,0}) = 0$; states $|1_{\sigma,0}\rangle$ have $n(E_\mathrm{A,0}) = n(- E_\mathrm{A,0}) = (1 + \sigma)/2$, where $\sigma = 1$ and $\sigma = -1$ correspond to spin up and down, respectively; finally, in the state $|2_0\rangle$ we have $n(-E_\mathrm{A,0}) = 0$ and $n(E_\mathrm{A,0}) = 1$. Ultimately, we obtain
\begin{equation}
\label{eq:kubo-vs-GRA-and-n}
\chi_{AB}^\mathrm{K}[\omega, n] = \int_{-\infty}^{\infty}d\varepsilon_{1}d\varepsilon_{2}\frac{n(\varepsilon_{1})-n(\varepsilon_{2})}{\hbar\omega+\varepsilon_{1}-\varepsilon_{2}+i0}\mathrm{Tr}\left[\mathcal{A}\mathcal{V}(\varepsilon_2)\mathcal{B}\mathcal{V}(\varepsilon_1)\right],\quad \mathcal{V}(\varepsilon)=\frac{i}{2\pi}(\mathcal{G}^{R}(\varepsilon)-\mathcal{G}^{A}(\varepsilon)).
\end{equation}
Notice that the trace in Eq.~\eqref{eq:kubo-vs-GRA-and-n} does not depend on the distribution function $n(\varepsilon)$ since it contains only the retarded and advanced Green's functions. The components of these Green's functions were computed in Appendix \ref{app:gf}. We identify
\begin{equation}
\label{eq:GR-components}
\mathcal{G}^{R/A}(\varepsilon) =
\begin{pmatrix}
G^{R/A}_{dd}(\varepsilon) & G_{dL}^{R/A}(\varepsilon) & G_{dR}^{R/A}(\varepsilon)\\
G_{Ld}^{R/A}(\varepsilon) & G_{LL}^{R/A}(\varepsilon) & G_{LR}^{R/A}(\varepsilon)\\
G_{Rd}^{R/A}(\varepsilon) & G_{RL}^{R/A}(\varepsilon) & G_{RR}^{R/A}(\varepsilon)
\end{pmatrix},
\end{equation} where $G_{dd}^{R/A}$ is determined by Eq.~\eqref{eq:G-inv} (with $\varepsilon$ exchanged for $\varepsilon \pm i0$), while $G^{R/A}_{id}$, $G^{R/A}_{di}$, and $G^{R/A}_{ij}$ are given in Eqs.~\eqref{eq:id_app} and \eqref{eq:exact-greens}. Note that the advanced Green's function can be obtained from the retarded Green's function via $\mathcal{G}^A(\varepsilon) = (\mathcal{G}^R(\varepsilon))^\dagger$. Matrix $\mathcal{V}$ can be decomposed into a continuum contribution (that is non-zero for $|\varepsilon|>\Delta$ only) and the contributions corresponding to the ABS (which are non-zero only at $\varepsilon = \pm E_{\mathrm{A,0}}$),
\begin{equation}
    \label{eq:poles}
    \mathcal{V}(\varepsilon)=\mathcal{V}(\varepsilon)\theta(|\varepsilon|-\Delta)  + \mathcal{V}_+ \delta(\varepsilon - E_\mathrm{A,0}) + \mathcal{V}_- \delta(\varepsilon + E_\mathrm{A,0}).
\end{equation}
In this expression,
\begin{equation}
\mathcal{V}_{\pm} =
\begin{pmatrix}
\nu_{dd,\pm}    & \nu_{dL, \pm}  & \nu_{dR, \pm}\\
\nu_{Ld, \pm} & \nu_{LL, \pm} & \nu_{LR, \pm}\\
\nu_{Rd, \pm} & \nu_{RL, \pm} & \nu_{RR, \pm} 
\end{pmatrix},
\end{equation}
where
\begin{equation}
\nu_{dd,\pm} = \pm\frac{(\Delta^{2}-E_{\mathrm{A,0}}^{2})}{2\Gamma_{R}\Gamma_{L}\Delta^{2}}\frac{dE_\mathrm{A,0}}{d\cos\varphi}
\left(\begin{array}{cc}
\pm\frac{E_{\mathrm{A,0}}}{Z(E_{\mathrm{A,0}})}+\epsilon_{g} & \frac{\Delta}{\sqrt{\Delta^{2}-E_{\mathrm{A,0}}^{2}}}\sum_{i} \Gamma_i e^{i\varphi_i}\\
\frac{\Delta}{\sqrt{\Delta^{2}-E_{\mathrm{A,0}}^{2}}}\sum_{i} \Gamma_i e^{-i\varphi_i} & \pm\frac{E_{\mathrm{A,0}}}{Z(E_{\mathrm{A,0}})}-\epsilon_{g}
\end{array}\right),\quad \frac{1}{Z(\varepsilon)} = 1+\frac{\Gamma}{\sqrt{\Delta^2-\varepsilon^2}},
\end{equation}
and
\begin{equation}
\nu_{id,\pm} = g_{\pm E_\mathrm{A,0}} T_i \nu_{dd,\pm} ,\quad\nu_{di,\pm} = \nu_{dd,\pm} T_i^\dagger g_{\pm E_\mathrm{A,0}},\quad\nu_{ij} =  g_{\pm E_\mathrm{A,0}} T_i\nu_{dd,\pm} T_j^\dagger g_{\pm E_\mathrm{A,0}}.
\end{equation}
Equation~\eqref{eq:kubo-vs-GRA-and-n} combined with the representation Eq.~\eqref{eq:poles} allow us to break the response function into physically distinct contributions. Using the particle-hole symmetry and Eq.~\eqref{eq:poles} we obtain
\begin{gather}
\chi_{AB}^\mathrm{K}[\omega, n] =  -\int_{\Delta}^{+\infty}d\varepsilon_{1}d\varepsilon_{2}\frac{1}{\hbar\omega+\varepsilon_{1}+\varepsilon_{2}+i0}\mathrm{Tr}\left[\mathcal{A}\nu(-\varepsilon_{2})\mathcal{B}\nu(\varepsilon_{1})\right]-\notag\\
\label{eq:contributions}
-\int_{\Delta}^{+\infty}d\varepsilon_1\frac{1+n(E_{\rm A,0})-n(-E_{\rm A,0})}{\varepsilon_{1}-E_{\rm A,0}+\hbar\omega+i0}\mathrm{Tr}\left[\mathcal{A}\nu_+\mathcal{B}\nu(\varepsilon_{1})\right] -\int_{\Delta}^{+\infty}d\varepsilon_1\frac{1+n(-E_{\rm A,0})-n(E_{\rm A,0})}{\varepsilon_{1}+E_{\rm A,0}+\hbar\omega+i0}\mathrm{Tr}\left[\mathcal{A}\nu_-\mathcal{B}\nu(\varepsilon_{1})\right]-\\
-\left[n(-E_{\rm A,0})-n(E_{\rm A,0}))\right]\frac{\mathrm{Tr}\left[\mathcal{A}\nu_-\mathcal{B}\nu_+\right]}{2E_{\rm A,0}+\hbar\omega+i0}+c.c.(-\omega),\notag
\end{gather}
where $c.c.(-\omega)$ denotes the complex conjugate of all of the preceding terms in the equation, in which we also change $\omega \rightarrow - \omega$. Different terms in Eq.~\eqref{eq:contributions} correspond to different transition processes. 
The first line corresponds to a process in which a Cooper pair in the condensate is broken to produce two quasiparticle excitations in the continuum. The second line contains processes that involve both the ABS and the quasiparticle continuum. Finally, the first term in the third line corresponds to transition processes that involve only the ABS. The contributions to $\chi^{\rm K}_{AB}[\omega, n]$ stemming from the first two lines are small at small frequencies due to the large energy denominators. They can be neglected for $\hbar\omega \ll \Delta - E_\mathrm{A,0}$ in comparison with either the first term in the third line of \eqref{eq:contributions}, or with the adiabatic contribution to the response function. In the main text we assumed $E_\mathrm{A,0} \lesssim \Delta$ and thus it was enough to require $\hbar\omega \ll \Delta$ to neglect the first two lines of Eq.~\eqref{eq:contributions}.

Equation~\eqref{eq:contributions} can be used to obtain an especially simple expression for the diagonal components of the dynamic part of the response function, $\delta\chi_{AA}$. We find
\begin{gather}
\notag
\delta \chi_{AA}[\omega, n] = -\int_{\Delta}^{+\infty}\frac{d\varepsilon_{1}d\varepsilon_{2}}{\varepsilon_1 + \varepsilon_2}\frac{2\hbar^2\omega^2}{(\varepsilon_{1}+\varepsilon_{2})^2-(\hbar\omega+i0)^2}\mathrm{Tr}\left[{\cal A}\nu(-\varepsilon_{2}){\cal A}\nu(\varepsilon_{1})\right]-\\
-\sum_{\sigma=\pm}\left(1+\sigma\left[n(E_{\rm A,0})-n(-E_{\rm A,0})\right]\right)\int_{\Delta}^{+\infty}\frac{d\varepsilon_1}{\varepsilon_1 - \sigma E_{\mathrm{A,0}}}\frac{2\hbar^2\omega^2}{(\varepsilon_{1}-\sigma E_{\rm A,0})^2-(\hbar\omega+i0)^2}\mathrm{Tr}\left[{\cal A}\nu_{\sigma}{\cal A}\nu(\varepsilon_{1})\right]-\\
\notag
-\left[n(-E_{\rm A,0})-n(E_{\rm A,0}))\right]\frac{1}{2E_{\mathrm{A,0}}}\frac{2\hbar^2\omega^2}{4E_\mathrm{A,0}^2 - (\hbar\omega+i0)^2}\mathrm{Tr}\left[{\cal A}\nu_-{\cal A}\nu_+\right].
\end{gather}
Notice that the diagonal parts of the response function scale as $\propto \omega^2$ at small frequencies.

\subsection{Occupation rule in the absence of interactions}
In the absence of interaction the response functions satisfy the occupation rule,
\begin{equation}
    \label{eq:occ-rule-app}
    \frac{1}{2}\left(\chi_{AB}[\omega, 0] + \chi_{AB}[\omega, 2]\right) = \chi_{AB}[\omega, 1].
\end{equation}
This can be easily checked using Eq.~\eqref{eq:contributions} for the finite-frequency response function $\chi_{AB}[\omega, n]$. To do that, first note that $n(-E_\mathrm{A,0})-n(E_\mathrm{A,0}) = 1$ in the state $|0_0\rangle$ and $n(-E_\mathrm{A,0})-n(E_\mathrm{A,0}) = -1$ in the state $|2_0\rangle$. Therefore, the average of $n(-E_\mathrm{A,0})-n(E_\mathrm{A,0})$ between the even states is zero. At the same time, $n(-E_\mathrm{A,0}) - n(E_\mathrm{A,0}) = 0$ in the odd states. Thus Eq.~\eqref{eq:occ-rule-app} holds.

We believe that the validity of the occupation rule Eq.~\eqref{eq:occ-rule-app} in the absence of Coulomb interaction is not restricted to a particular model considered here. First of all, Eq.~\eqref{eq:occ-rule-app} is agnostic to the presence of other ABS within the gap, as long as they have a consistent occupation. Equation~\eqref{eq:occ-rule-app} should also hold at $U=0$ in the presence of magnetic field and spin-orbit coupling (though in the absence of spin-degeneracy the right hand side of Eq.~\eqref{eq:occ-rule-app} should be replaced with a half-sum of the odd sates in the considered spin-split doublet). Finally, the occupation rule also survives the presence of above-the-gap quasiparticles provided that the occupation of the ABS is certain.

\section{Effects of capacitance between the dot and the leads\label{app:cap}}
Throughout the main text, we assumed that the capacitance between the dot in the weak link and the gate, $C_g$, is much larger than capacitances between the dot and the leads, $C_L$ and $C_R$ (indices $L/R$ correspond to left and right lead, respectively). This may not necessarily be the case in the experiments. In this Appendix, we discuss the modifications of our theory that arise when capacitances between the leads and the dot are comparable to the capacitance between the gate and the dot. We still neglect the capacitance between the dot and the ground, assuming that the grounded parts of the device are located sufficiently far away from the weak link.

The main effect of appreciable $C_R$ and $C_L$ is the modification of a relation between the energy of the level at the dot, $\epsilon_g(t)$, and the voltages $V_g(t)$, $V_L(t)$, and $V_R(t)$ [here $V_g(t)$ is the voltage bias applied to the gate, $V_L(t)$ is the voltage at the left lead, and $V_R(t)$ is the voltage at the right lead]. We find
\begin{equation}
    \label{eq:epsilon-cap}
    \epsilon_g(t) = -e\frac{C_g V_g(t) + C_L V_L(t) + C_R V_R(t)}{C_g + C_L + C_R }.
\end{equation}
This expression should be contrasted with a simpler relation used in the main text, $\epsilon_g(t) = -eV_g(t)$; the latter relation is justified only when $C_g \gg C_L,C_R$. Keeping in mind the modified expression for $\epsilon_g(t)$, Eq.~\eqref{eq:epsilon-cap}, the low-energy Hamiltonian is still given by $H_\mathrm{even}^{\mathrm{(le)}} = E_\mathrm{even} + \mathcal{H} + \Theta$, where $E_\mathrm{even}$ is determined by Eq.~\eqref{eq:even}, $\mathcal{H}$ is determined by Eq.~\eqref{eq:low-en-ham_int_app}, and $\Theta$ is determined by Eq.~\eqref{eq:theta_int_app}.

Next, we derive the relation between the admittance matrix of the weak link and the response functions $\chi_{II}$, $\chi_{IQ}$, $\chi_{QI}$, and $\chi_{QQ}$, taking into the account $C_L$ and $C_R$. Motivated by cQED applications of Section \ref{sec:cqed}, we assume that the gate voltage is static. Then we find  
\begin{subequations}
\label{eq:admit}
    \begin{align}
        Y_{LL}[\omega]&=
        \frac{\chi_{II}[\omega]}{-i\omega}+\frac{C_g-C_L+C_R}{C_g+C_L+C_R}\frac{\chi_{IQ}-\chi_{QI}}{2}+\left(\frac{C_g - C_L + C_R}{C_g + C_L + C_R}\right)^2\frac{i\omega\chi_{QQ}}{4},\\
        Y_{LR}[\omega]&=
        \frac{\chi_{II}[\omega]}{i\omega}-\frac{1}{2}\frac{C_g(\chi_{IQ}+\chi_{QI})-(C_L-C_R)(\chi_{IQ}-\chi_{QI})}{C_g+C_L+C_R}+\frac{C_g^2-(C_L-C_R)^2}{(C_g+C_L+C_R)^2}\frac{i\omega\chi_{QQ}}{4},\\
        Y_{RL}[\omega]&=
        \frac{\chi_{II}[\omega]}{i\omega}+\frac{1}{2}\frac{C_g(\chi_{IQ}+\chi_{QI})+(C_L-C_R)(\chi_{IQ}-\chi_{QI})}{C_g+C_L+C_R}+\frac{C_g^2-(C_L-C_R)^2}{(C_g+C_L+C_R)^2}\frac{i\omega\chi_{QQ}}{4},\\
        Y_{RR}[\omega]&=
        \frac{\chi_{II}[\omega]}{-i\omega}-\frac{C_g+C_L-C_R}{C_g+C_L+C_R}\frac{\chi_{IQ}-\chi_{QI}}{2}+\left(\frac{C_g + C_L - C_R}{C_g + C_L + C_R}\right)^2\frac{i\omega\chi_{QQ}}{4}.
    \end{align}
\end{subequations}
Matrix $Y_{ij}[\omega]$ can be used to determine the frequency shift of the microwave resonator coupled to the weak link, see Eq.~\eqref{eq:dw}.
\end{document}